\documentclass{aastex}

\usepackage[]{emulateapj5}

\usepackage{graphicx}

\slugcomment{To be published in the ApJ %, draft: \today
}

\shorttitle{Deep ACS Imaging of the Halo of NGC 5128}
\shortauthors{Rejkuba et al.}

\begin{document}

\title{Deep ACS Imaging of the Halo of NGC 5128: Reaching the Horizontal
	Branch\thanks{Based on observations made with the NASA/ESA Hubble
	Space Telescope, obtained at the Space Telescope
	Science Institute, which is operated by the Association of Universities for
	Research in Astronomy, Inc., under NASA contract NAS 5-26555. These
	observations are associated with program \#GO-9373.
	Also partially based on observations collected at the European Southern 
    	Observatory, La Silla, Chile, within the Observing Programme
	071.D-0560.}}

\author{Marina Rejkuba}
\affil{European Southern Observatory, Karl-Schwartzschild-Strasse
           2, D-85748 Garching, Germany}
\email{mrejkuba@eso.org}

\author{Laura Greggio}
\affil{INAF, Osservatorio Astronomico di Padova, Vicolo dell'Osservatorio 5,
35122 Padova, Italy}
\email{greggio@pd.astro.it}

\author{William E.\ Harris}
\affil{Department of Physics and Astronomy, McMaster University, Hamilton ON
L8S 4M1, Canada}
\email{harris@physics.mcmaster.ca}

\author{Gretchen L.\ H.\ Harris}
\affil{Department of Physics, University of Waterloo, Waterloo ON N2L 3G1, Canada}
\email{glharris@astro.uwaterloo.ca}
\and

\author{Eric W.\ Peng\altaffilmark{2}}
\affil{Department of Physics and Astronomy, Rutgers University, Piscataway, NJ
08854, USA}
\email{Eric.Peng@nrc-cnrc.gc.ca}

\altaffiltext{2}{present address: Herzberg Institute of Astrophysics, 
National Research Council, 5071 West Saanich Road, Victoria, BC V9E 2E7, Canada}

\begin{abstract}
Using the HST Wide Field Camera (WFC) of the Advanced Camera for Surveys (ACS),
we have obtained deep $(V,I)$
photometry of an outer-halo field in NGC~5128, to a limiting magnitude of 
$I\simeq 29$.
Our photometry directly reveals the core helium-burning stellar
population (the ``red clump'' or horizontal branch) in a giant E/S0
galaxy for the first time.  

The color-magnitude diagram displays a very wide red 
giant branch (RGB), an asymptotic giant branch (AGB) bump, and the 
red clump; no noticeable population of blue HB stars is present,
confirming previous suggestions that old, very metal-poor population
is not ubiquitous in the halo of this galaxy.
From the upper RGB we derive the metallicity distribution, which we find
to be very broad and moderately metal-rich, with average 
$\mathrm{[M/H]}=-0.64$ and dispersion $0.49$~dex. 
The MDF is virtually identical to that found in other halo fields observed 
previously with the HST, but with an enhanced metal-rich population
which was partially missed in the previous surveys due to $V$-band
incompleteness for these very red stars. Combining the 
metallicity sensitive colors of the RGB stars with the metallicity
and age sensitive features of the AGB bump and the red clump, we infer the
average age of the halo stars to be $\sim 8^{+3}_{-3.5}$~Gy. 

As part of our study, we present an empirical calibration of the 
ACS $F606W$ and $F814W$
filters to the standard $V$ and $I$ bands, achieved with ground-based
observations of the same field made from the EMMI camera of
the New Technology Telescope of the ESO La Silla Observatory.

\end{abstract}


%
%-------------------------------------------------------------------
%

\section{Introduction}

A significant fraction of all the stars in the Universe reside within giant
elliptical galaxies, yet the formation epochs of these large systems 
are still
a matter of debate. The three main elliptical galaxy formation processes are generally
considered to be: (i) {\it in situ} conversion of protogalactic gas through
multiple starbursts \citep[e.g.][]{partridge&peebles67,tinsley72,larson74,
harris+95}, (ii) later mergers of pre-existing disk
galaxies \citep[e.g.][]{toomre77,ashman&zepf92,kauffmann+93,cole+00}, 
and (iii) ongoing accretion of small satellites \citep[e.g.][]{cote+98}. 
The three models predict different mean ages; they also predict 
different metallicities and metallicity
gradients for the stars that build these galaxies, as well as 
different distributions and characteristics of the (multiple) globular cluster
populations. 

Elliptical galaxy formation studies are generally 
based on observations of high-redshift populations, or deductions of
mean ages and metallicities of nearby systems from measurements of 
their integrated spectral indices \citep[e.g.][]{trager+00, 
kuntschner+02, dressler+04}. 
In recent years the availability of 
high-multiplex instruments at large aperture telescopes makes the study of
globular clusters and planetary nebulae in these galaxies a promising 
new tool \citep[e.g.][]{walsh+99, romanowsky+03, puzia+04, peng+04a, peng+04b}.   
At the same time, it has now become possible to
study the {\sl resolved stellar populations} 
in elliptical galaxies directly --
an important complementary approach to the study of the high-redshift
Universe, and integrated stellar and globular cluster light.

But even with the best available imaging tools (HST, VLT, Gemini), 
resolving the old halo stars is possible for only the nearest 
E/S0 galaxies. Pioneering work of this type was done notably for the 
halos of large E/S0 systems including NGC~5128 \citep{soria+96}, 
Maffei~1 \citep{davidge+vandenbergh01}, NGC 3379 \citep{gregg+04}, 
NGC~3115, NGC~5102, and NGC~404 \citep{schulte-ladbeck+03}. Hence, 
these few objects provide the irreplaceable chance
for calibrating all the techniques that involve
integrated spectral indices and population synthesis. Moreover they 
provide the only way to measure the properties of stellar populations 
in the outer halo, while the integrated light measurements refer to 
the central areas.

At a distance of 3.8 Mpc NGC~5128 (commonly referred to as Centaurus A) 
is by far the nearest easily observable E/S0\footnote{sometimes also referred 
to as S0pec} galaxy 
\citep{harris+99, rejkuba04}, and is thus a uniquely valuable 
testing ground for stellar population and galaxy formation models.  
Its halo light was first resolved in stars by \citet{soria+96} using 
the HST/WFPC2 camera. Further, deeper WFPC2 photometry, reaching 
2.5 magnitudes down the red giant branch (RGB), was carried out by 
\citet{harris+99} and \citet{harris&harris00, harris&harris02} 
in the optical $V$ and $I$ bands, and by \citet{marleau+00} with 
the near-infrared NICMOS camera in $J$ and $H$. Using ground-based data, 
\citet{rejkuba+01,rejkuba+03}  presented deep color-magnitude 
diagrams (CMDs) in $V$, $J$, $H$, and $K_s$ with FORS1 and ISAAC at the VLT.

All these earlier studies indicated that the halo of this giant E/S0 galaxy is
dominated by the presence of normal RGB stars, demonstrating that the halo
stars are at least a few Gy old.  The same RGB photometry provides the 
means to construct the metallicity distribution function (MDF) of its halo
stars.  However, relatively little is yet known about the {\em age distribution 
function} (ADF) of its halo stars. There is ongoing star formation in the 
north-eastern halo fields along the galaxy's major axis 
\citep{graham98, mould+00, fg00, rejkuba+01, gf02, rejkuba+02}, 
but simulations show that the star formation rate in this area has been 
lower than in the solar neighborhood and comparable to some Local Group 
dIrr galaxies, hence contributing relatively little to the stellar halo
mass \citep{rejkuba+04}. A suggestion that this galaxy had an 
appreciable intermediate-age component was raised in the early work of 
\citet{soria+96} and \citet{marleau+00}, but was contested later with 
observations of different halo fields 
\citep{harris+99, harris&harris00, harris&harris02} which indicated no 
need for any significant contribution of intermediate-age or young 
components in the halo. However, the extended giant branch observed in 
the $V,K$ CMD \citep{rejkuba+01} and the presence of long-period variable 
AGB stars with periods in excess of 450-500 days \citep{rejkuba+03}, seem 
to confirm that up to $\sim 10$\% of the halo might consist of an 
intermediate-age population in the range $\sim 3-7$~Gy. Additional lines 
of evidence for an intermediate-age component come from recent photometry 
and spectroscopy of the NGC~5128 globular cluster 
system \citep{peng+04b, yi+04}.

The Advanced Camera for Surveys \citep[ACS;][]{ford+98} 
installed on HST during the
servicing mission 3B in March 2002 has made it possible to investigate
stellar populations to both significantly deeper limits and higher
resolution.  With ACS, it is now possible in NGC 5128 to reach the red 
clump and the horizontal branch (the old core-helium-burning stars) and 
the asymptotic giant branch (AGB) bump directly. Since these features are 
sensitive to both metallicity and age, we can examine quantitatively 
and directly the ADF of the halo stars in a giant E/S0 galaxy for the 
first time. In this paper we present a very deep color-magnitude
diagram (CMD) of a remote halo field in NGC 5128, showing the AGB bump, 
red clump and the horizontal branch in NGC~5128. Simple comparisons of 
the observed and simulated luminosity functions are used to obtain an 
average age for the halo stars. In a following paper we plan to present 
a more thorough simulation of the CMD and
derive from it a star formation history (SFH).

Throughout the paper we use an intrinsic distance modulus 
for NGC 5128 of $(m-M)_0 = 27.92$ 
\citep{rejkuba04} and a reddening of $E(B-V)=0.11$ \citep{burstein,schlegel}.

%
%--------------------------------------------------------------------
%

\section {The Data}
\label{data}

\subsection{HST Observations and Photometry}

For the deep HST ACS observations of the NGC~5128 halo we chose a field
centered at $\alpha_{\mathrm{J2000}}=13^{\mathrm h} 25^{\mathrm h} 15\fs1, 
\delta_{\mathrm{J2000}}=-43^\circ 34\arcmin 30\arcsec$, a projected distance of
some $33\farcm 3$ to the south of the galaxy center. This location corresponds
to a linear projected distance of 38 kpc.  The choice of 
such a remote halo location was driven by the requirements to have low surface
brightness, i.e. low enough stellar density to permit high precision photometry, 
unaffected by crowding, but still give a high enough number of
stars to make up a large statistical sample.  We also avoided, 
as much as possible, the obvious bright foreground stars which would 
saturate in
long exposures. This latter criterion proved to be an issue with the 
combination of the low galactic
latitude of NGC~5128 and the high sensitivity of the new camera. Additionally
the target field was chosen to avoid any peculiarities
of this galaxy, like jet induced star forming regions \citep{mould+00,
rejkuba+02}, shells \citep{malin+83}, and dust lanes \citep{stickel+04}.

\begin{figure}
\plotone{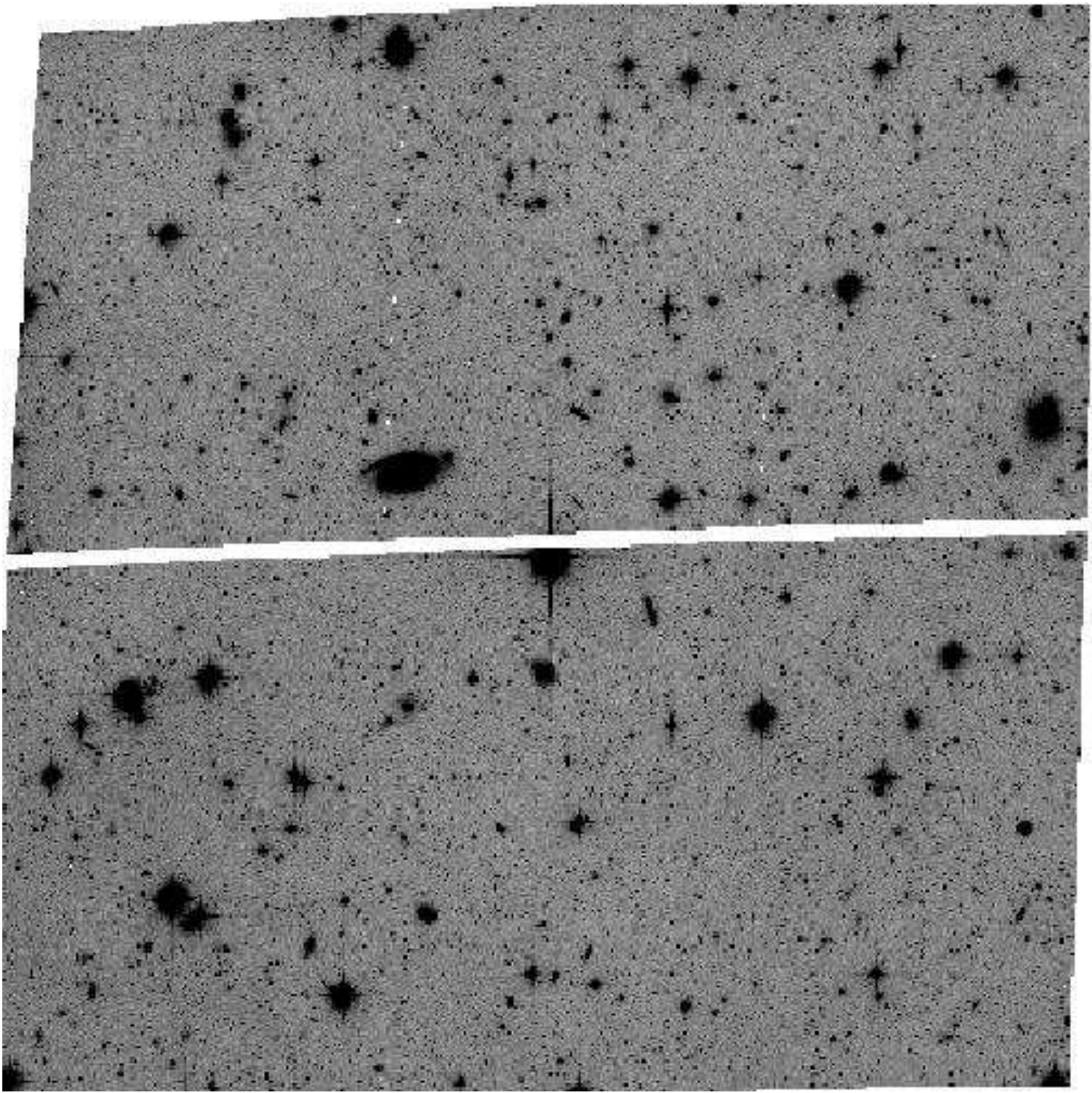}   
  \caption[]{Combined image consisting of 12 V and 12 I full-orbit
	exposures. The total exposure time is more than 19~h. 
	The faintest stars in this
	image have $V\sim30$ and $I\sim29$~mag. }
  \label{VIcomb}
\end{figure}

The ACS observations were taken in Cycle 11 between 2002 July 5 and 9 
within GO program 9373.  They consist of 12 full 
orbit exposures in F606W (broad $V$), divided among 4 visits, 
and 12 full orbit F814W ($I$) exposures, also divided among 4 visits. 
The first exposure in each visit was 2520 sec, while
all the others were 2600 sec long. Hence the total exposure times amount to
$4 \times 2520 + 8 \times 2600$ sec, or 8.58~h for each filter. The individual
exposures were not dithered, but the pointing differences between the
different visits resulted in shifts between images of up to 4 pix, which,
combined with 12 exposures per filter, enabled us to remove the cosmic rays and
hot pixels efficiently by registering and combining the individual frames.
The very deep median combined image made from all the exposures is shown in
Fig.~\ref{VIcomb}. 

The data were processed with the on-the-fly CALACS pipeline which 
included dark and bias subtraction, flatfielding and geometric distortion 
correction, but no cosmic ray rejection. 
We used the pipeline reduced ``drz'' images for photometry
with the following steps.
First, we multiplied each image by the corresponding exposure time. 
Then we ran standard PSF photometry with the DAOPHOT suite of 
codes \citep{stetson,stetsonALF}. 
The PSF for each individual exposure was determined with the spatially variable
option and $65-90$ non-saturated, relatively isolated stars 
that did not suffer from many cosmic ray hits. The PSFs from the
12 $F606W$ and 12 $F814W$ images were then combined into a single PSF per filter
using the MULTIPSF program. The FIND+PHOT+ALLSTAR programs were followed by 
DAOMATCH and DAOMASTER in order to determine accurate
geometric transformations between the images, and then all 
the images (both in F606W and F814W filters) were combined together 
with MONTAGE2. Using the resulting 
deep, cleaned image (Fig.~\ref{VIcomb}) we constructed
an initial photometry list of 70586 candidate stars, which was then used 
as a coordinate input for ALLFRAME. The resulting photometry for 65000 
stars in both images was used to determine new, improved geometric 
transformations between images, with the requirement to detect the
same star on all the images and adopting stringent photometric 
quality ($\sigma<0.05$). All of the 24 frames were then combined again and we 
made two runs of FIND+PHOT+ALLSTAR to define a final input list of 
93812 stars for ALLFRAME. The increase in the number of detected stars on the
combined image is due to much improved transformations which were based on more 
than 60000 stars compared to the few hundred that could be detected on 
individual images in the first run of ALLSTAR. 

\begin{figure}
\plottwo{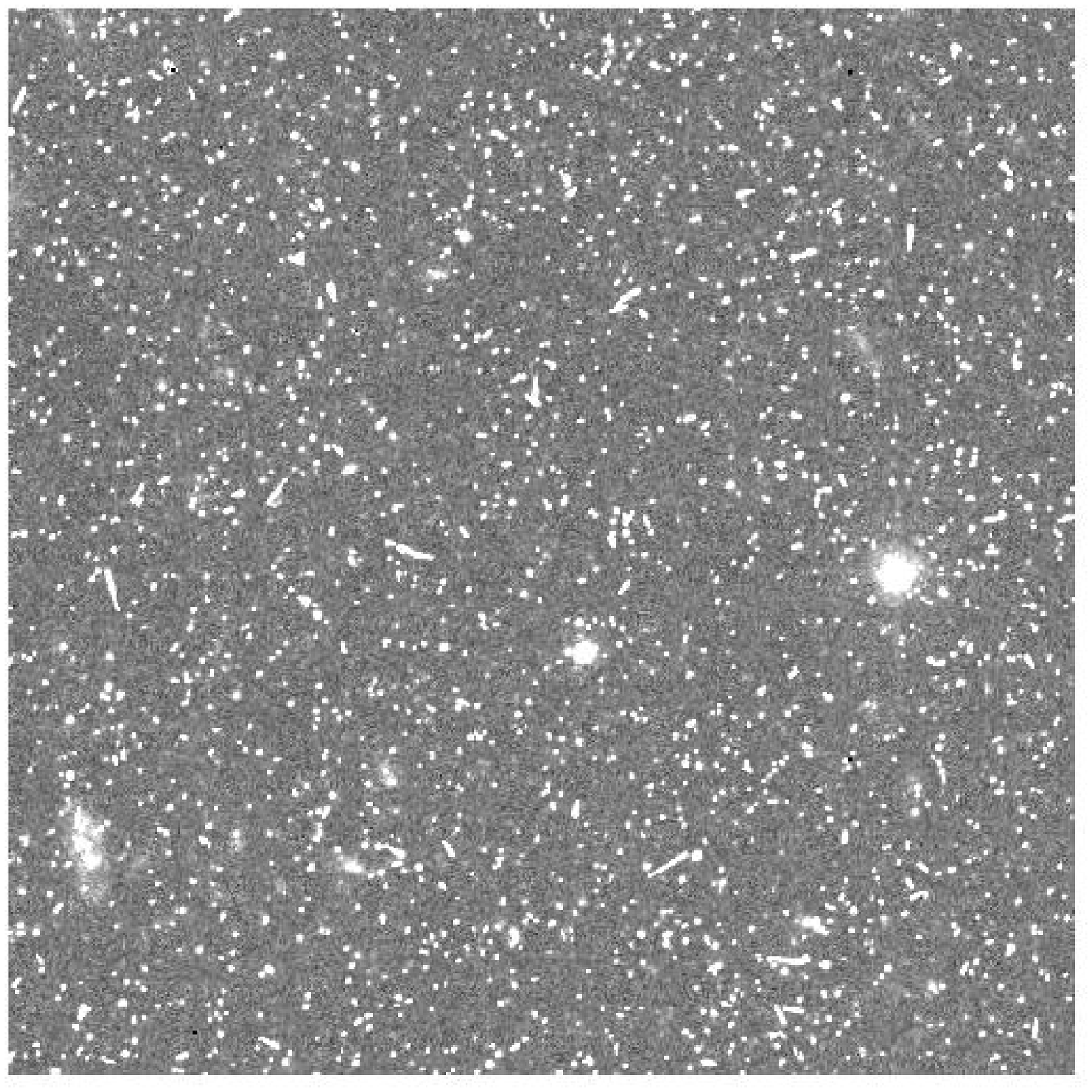}{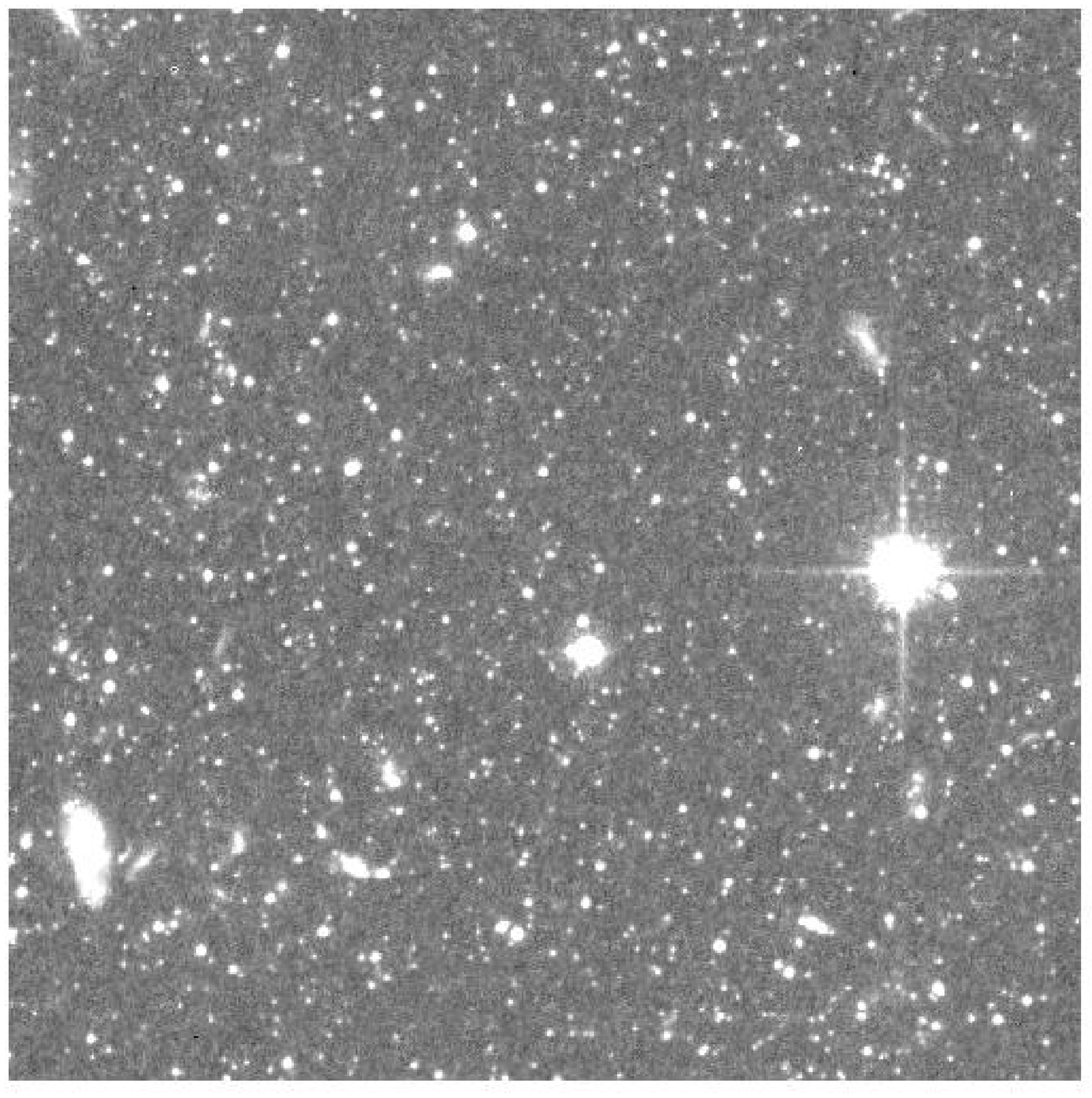}   
  \caption[]{Left panel: a $45'' \times 43''$ portion of one F606W image. 
  	Most of the sources visible are cosmic rays. Right panel: The same 
	part of the observed field as on the left is show this time as a 
	combined image consisting of all 12 F606W and 12 F814W exposures. 
	The total exposure time is more than 19~h. A large number of 
	background galaxies is visible. They are rejected from our final 
	photometry catalogue by very stringent 
	photometric selection criteria.} 
  \label{Vima}
\end{figure}

It should be noted that the long individual exposure times meant that 
all the images were affected by a very large number of cosmic rays. 
An example of a small portion of one $F606W$ exposure is shown in 
the left panel of Fig.~\ref{Vima}. The final combined image of the
same $45'' \times 43''$ field is in the right panel.
The cosmic rays were not removed prior to ALLFRAME PSF photometry. 
The stars were detected on the basis of their matched coordinates.
The final photometry catalogue contains 93306 F814W and 93067 F606W 
sources detected on at least 6 frames of each filter. The requirement 
to detect a star in at least 6 frames effectively rejects all the cosmic 
rays located near or on images of real stars in the field. The mean 
magnitude and the associated error were calculated for all the stars for 
each filter and then the two lists were matched with the requirement that 
the star is detected in both filters. 

\begin{figure}
\includegraphics[angle=270,width=15cm]{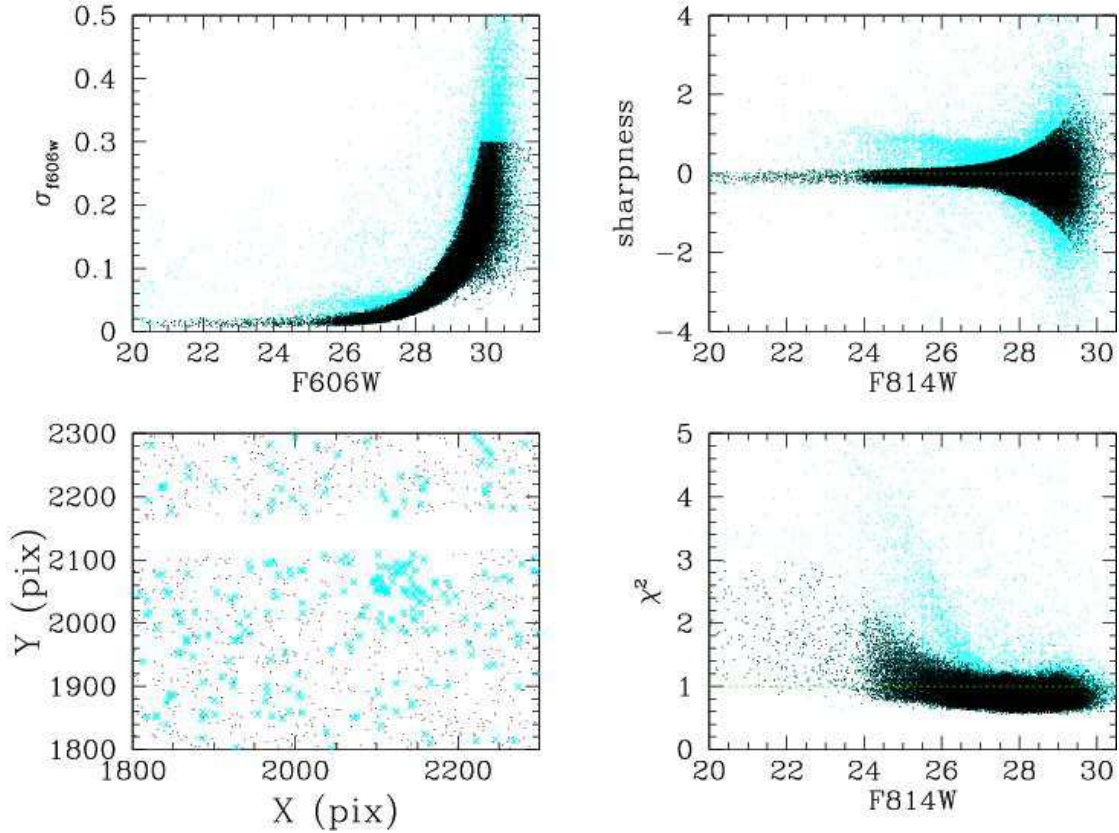}   
  \caption[]{Photometry selection criteria based on photometric error, 
  sharpness and $\chi^2$ parameters from DAOPHOT programs are plotted as 
  function of F606W and F814W magnitudes and positions on the chip. Only 
  sources lying within two hyperbolic envelopes around the {\em sharpness} 
  value of zero (upper right panel) and to the right of the maximum allowed 
  $\sigma$ at a given magnitude (upper left panel), as well as having 
  $\sigma<0.3$ and $\chi^2 < 3$ and $-2< sharpness < 2$ are considered 
  ``good'' stars. They are plotted with black dots in all the panels. 
  The sources that did not satisfy the above criteria are plotted with 
  light gray dots (cyan color in electronic edition) and were rejected 
  from the final photometric catalogue. In the lower left panel we show that 
  these selection criteria effectively reject all the noise spike detections 
  around a saturated star: a large number of gray (cyan) crosses around 
  (x, y)=(2100, 2080) corresponds to a very saturated foreground star.}
  \label{seleviall}
\end{figure}

Additional selection criteria were applied, as illustrated in 
Fig.~\ref{seleviall}, on DAOPHOT parameter {\em sharpness}
and on $\sigma$ (photometric error). The {\em sharpness} parameter 
estimates the intrinsic  angular size of the measured  objects. It is 
roughly defined as the difference between the square of the width of the object 
and the square  of  the  width  of PSF and has values close to zero for single 
stars, large positive values for blended doubles and partially resolved galaxies,
and large negative values for cosmic rays and blemishes.
We chose as ``good'' stars those 
sources lying within two hyperbolic envelopes around the {\em sharpness} 
value of zero (upper right panel of Fig.~\ref{seleviall}) and to the right
of the maximum allowed $\sigma$ at a given magnitude (upper left panel). 
The remaining cosmic ray blemishes and resolved background galaxies
were then rejected. Additionally we restricted the maximum allowed 
absolute value of the {\em sharpness} parameter to 2 (upper right panel), 
of $\sigma$ to 0.3 (upper left panel) and of $\chi^2$ to  
3 (lower right panel). In the lower left panel of Fig.~\ref{seleviall} 
we show that these selection criteria effectively reject all the noise spike 
detections around a saturated star (a large number of gray crosses around 
(x, y)=(2100, 2080) corresponds to a severely saturated foreground star). Also, 
it can be seen that the crowding is not excessive, hence allowing good sky 
subtraction and reliable photometry. These final selections resulted in a 
total of 77810 objects detected and measured in the entire field.

\begin{figure}
\includegraphics[angle=270,width=15cm]{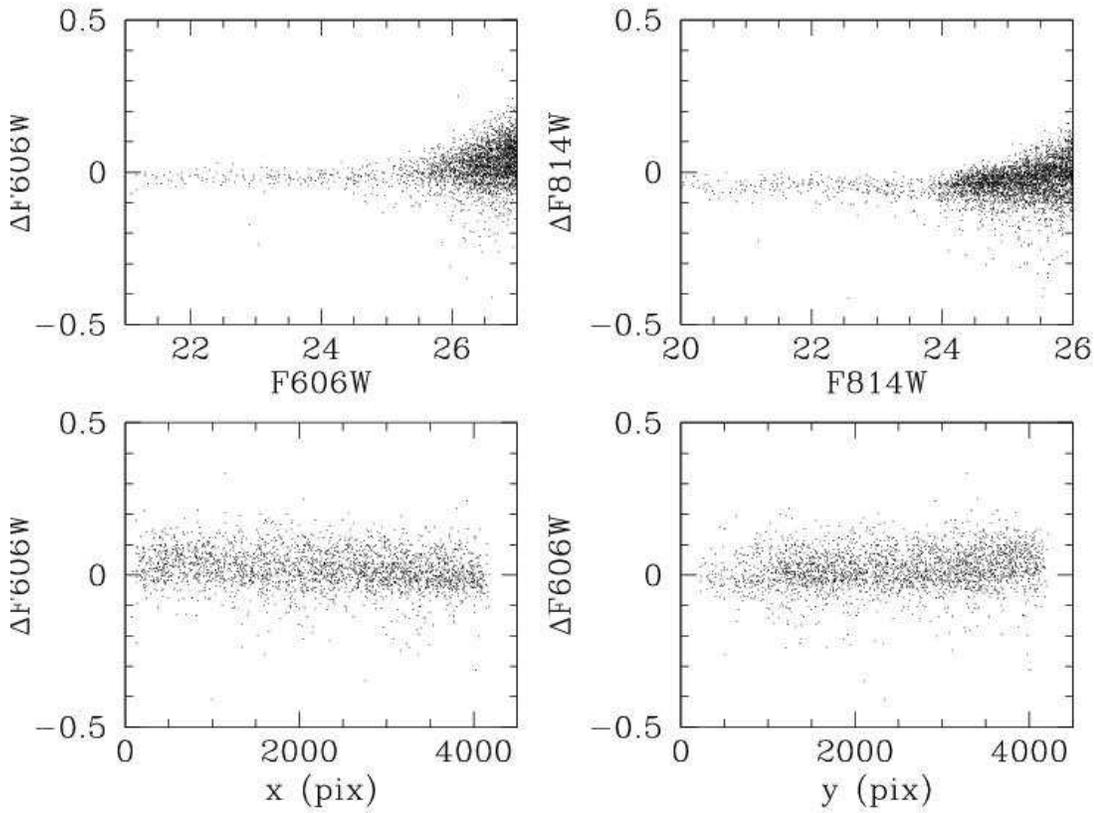}   
  \caption[]{Comparison between our final photometric catalogue
  and the photometry done on a subset of flatfielded, but not drizzled images.
  }
  \label{flt_drz}
\end{figure}

As suggested by the referee, we have checked our photometry against a reduction
based on a subset of flatfielded, but not drizzled images. The reduction of
``flt'' images was done as described in \citet{bedin+05}, and
the comparison is shown in Fig.~\ref{flt_drz}. The small offset  
between the measurement done on ``flt'' and ``drz'' images may come from the 
difference in the aperture correction. Any systematic effects are smaller than 
0.01 mag. The larger scatter for 
fainter magnitudes is due to smaller number of ``flt'' images. This comparison
shows that the process of drizzling, which results in adjacent pixels being
correlated, has not affected the precision of our photometry.

\begin{figure}
\includegraphics[width=8cm]{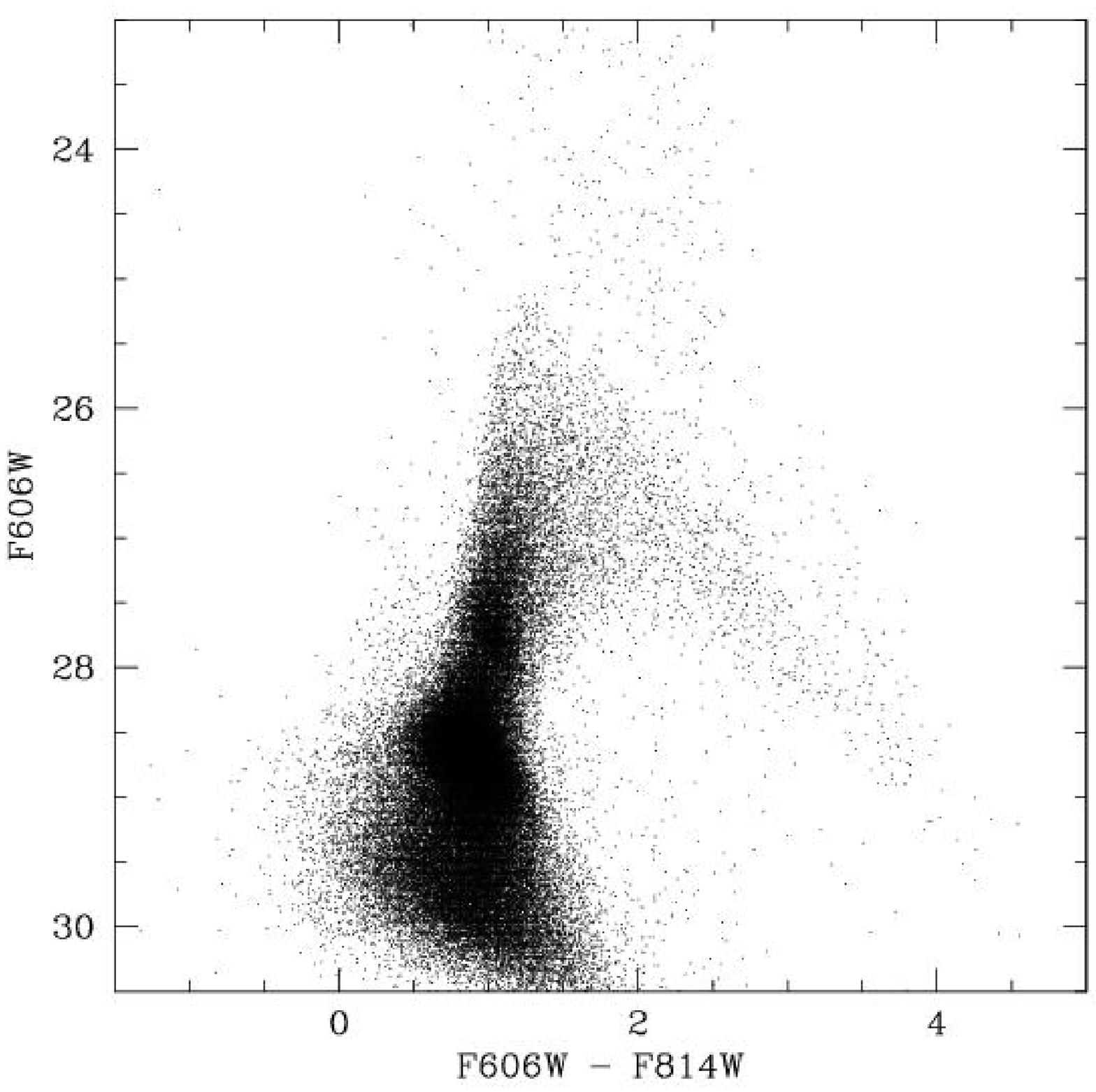}   
\includegraphics[width=8cm]{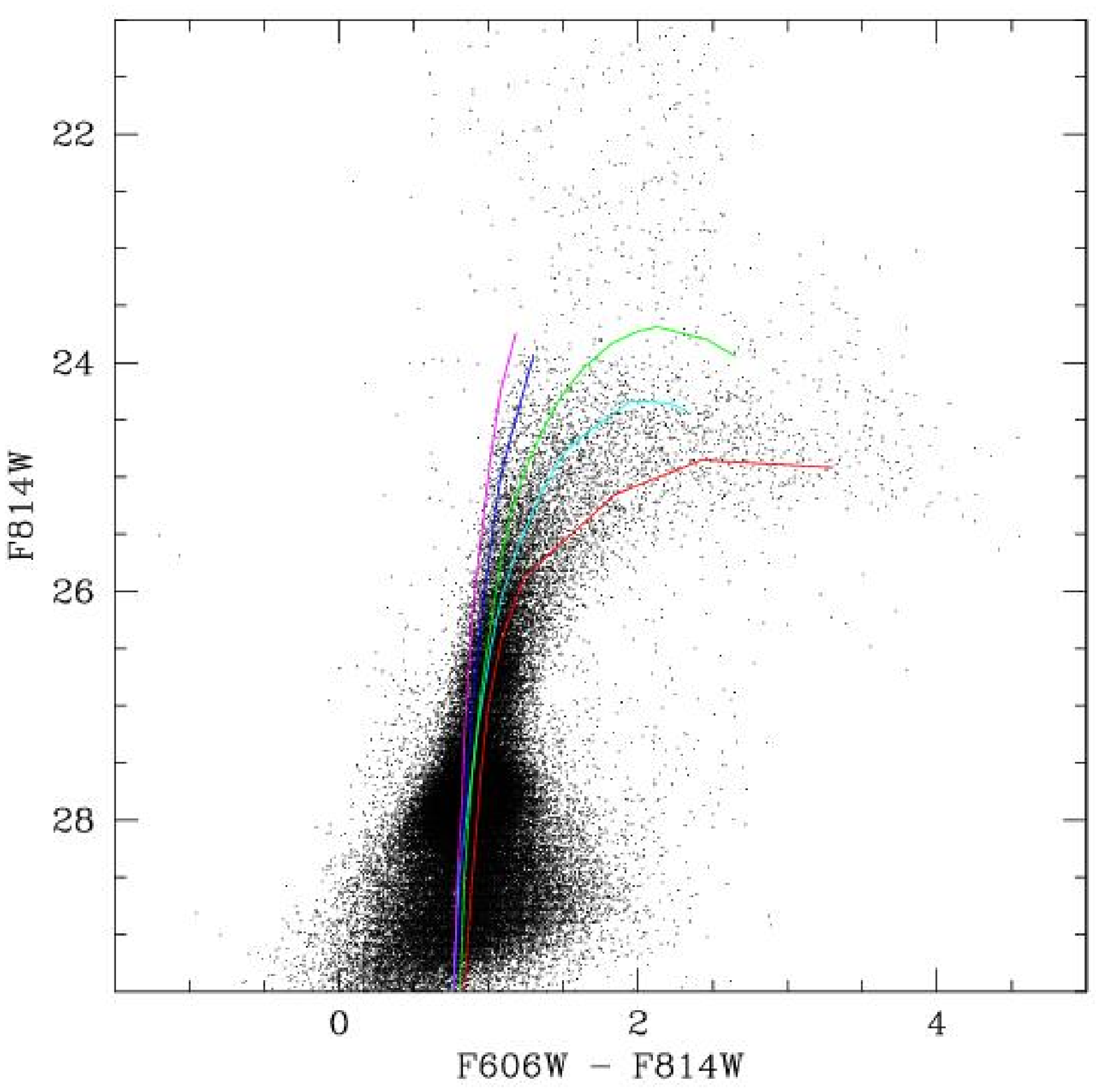}   
  \caption[]{Left panel: F606W-F814W vs. F606W CMD calibrated to VEGAmag HST 
  system. Right panel: F606W-F814W vs. F814W CMD calibrated to VEGAmag HST 
  system. Overplotted are RGB fiducials derived from the ACS observations 
  with the same filters by \citep{bedin+05}.}
  \label{CMD_VEGAMAG}
\end{figure}

To calibrate the photometry to the VEGAmag HST photometric system we adopted
the zeropoints from the ACS web page (December 2003), namely 26.385~mag for 
F606W and 25.487~mag for F814W.  An aperture correction
of -0.09 mag in both filters was adopted to scale magnitudes from 
$0\farcs 5$ to an infinite radius (M.\ Sirianni, private communication). 
The CMDs calibrated to the VEGAmag system are presented in 
Fig.~\ref{CMD_VEGAMAG}. Overplotted on the $F814W~vs.~(F606W-F814W)$ CMD 
are RGB fiducials from \citet{bedin+05} 
derived from fits to the Galactic globular clusters 
NGC~6341, NGC~6752, NGC~104 (47~Tuc), NGC~5927, and  NGC~6528 
observed with ACS in the same photometric bands \citep{brown+03}. 
These clusters span a range of metal abundances 
from $-2.2$~dex up to Solar metallicity. Already in this figure 
the large range
of metal-abundances among the halo stars in NGC~5128 can be appreciated.
We note that the RGB ridge line for NGC~104, which has intermediate metallicity
between metal-poor (NGC~6341 and NGC~6752) and metal-rich clusters
(NGC~5927 and  NGC~6528) appears to extend to much brighter magnitudes than
other clusters. This is not the case when the comparison is made in $V$ and 
$I$ bands and deserves clarification, which is beyond the scope of the present
paper. The calibration of the photometry 
to the standard ground based $V$ and $I$-bands is described below.

\subsection{Ground Based Observations and Photometry}

For the purposes of obtaining an empirical calibration of the ACS photometry 
to the ground based system, on Aug.\ 1 2003, we obtained $5 \times 330$~sec 
$V-$band exposures and $4 \times 320$~sec $I-$band exposures in our ACS 
target field using the EMMI optical imager at the New Technology Telescope
(NTT) at the ESO La Silla Observatory \citep{emmi}. 
We used the red arm of EMMI (RILD),
which has a $9.0' \times 9.9'$ field of view; we also used $2 \times 2$ 
binning giving an image scale $0\farcs33 / \mathrm{pix}$. The weather was 
photometric and seeing was 1.2 arcsec.

The reduction procedure for the EMMI data involved bias
subtraction and division by normalized twilight flat field images.
Individual dithered exposures taken with the same filter 
were then shifted and combined into a single deep image in each filter on which 
photometry was performed with DAOPHOT \citep{stetson}.  A spatially variable
PSF was constructed for both images using more than 60 stars. Only one cycle of 
FIND+PHOT+ALLSTAR was run as the fainter stars did not have
sufficient photometric accuracy for calibration. 

The calibration of the EMMI photometry was established from 57 stars in 3 
Landolt fields \citep{landolt92, stetson00}  PG1525, PG2331, and SA~111, 
all observed on the same night. The following photometric solution was derived:
\begin{eqnarray}
v& =& V  - 26.002 (\pm 0.043) + 0.125 (\pm 0.036) * Airmass\\
i& =& I  - 25.614 (\pm 0.046) + 0.054 (\pm 0.035) * Airmass\cr
 &  &  + 0.040 (\pm 0.011) * (V-I)
\end{eqnarray}
where $v$ and $i$ are instrumental magnitudes and $V$ and $I$ are magnitudes from
the \citet{stetson00} catalogue. 
The $1\sigma$ scatter around the mean is 0.034 mag for
both filters. Including the color term in the $V$-band equation does not reduce 
the scatter, and the error in the color term is larger than its value, 
hence we prefer not to include it.

\section{Calibration of the ACS $F606W$ and $F814W$ photometry}

Most of the stars observed with EMMI belong to the foreground Galactic population. 
Given the exposure times, these stars are 
quite bright and some are saturated on the ACS images. 
The match between the ACS and EMMI catalogs was obtained with the
IRAF\footnote{The Image Reduction and Analysis Facility (IRAF) is distributed by
the National Optical Astronomy Observatories, which are operated by the
Association of Universities for Research in Astronomy, Inc., under contract with
the National Science Foundation} task
GEOMAP and was further refined with DAOMASTER \citep{stetson}. 
The matched sources in both images number just 119 and each one of them was 
visually inspected. This resulted in rejection of 52 matched sources, 
either because they were extended or
double on the ACS and single on EMMI images, 
or because they were saturated in the ACS image. 
%In a few cases 
%where the count levels of a star in the ACS image exceeded the saturation 
%level, but the profile of the star looked
%smooth and followed the non-saturated star PSF, that star was retained. 
%The list of all common sources is listed in Tab.~\ref{EMMI_ACS.tab}. 
%The comment in the last
%column indicates the reason for rejecting the star from the fit.

\begin{figure}
\includegraphics[angle=270,width=16cm]{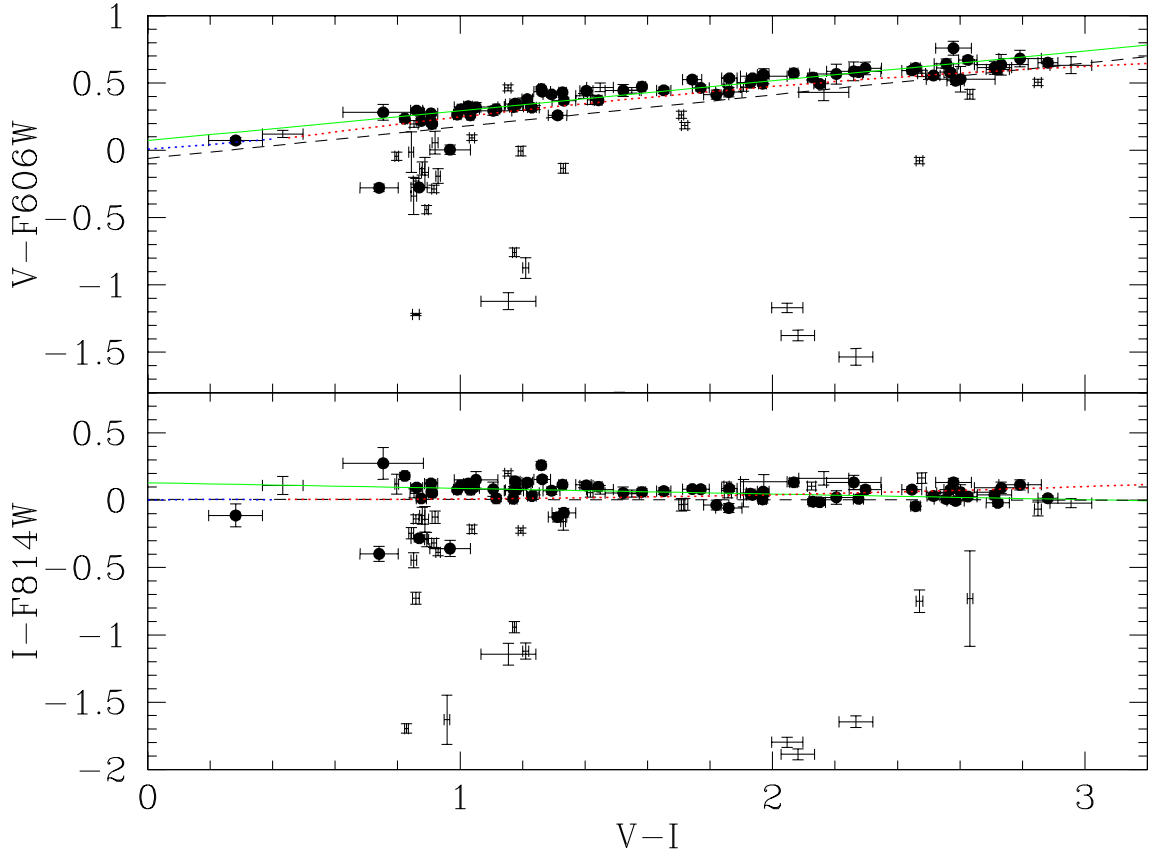}
  \caption[]{Empirical calibration of the ACS photometric system 
  (F606W and F814W) into the $V$ and $I$ bands. All the 119 matching 
  sources between the ACS and EMMI images are plotted with error-bars, and 
  67 stars used to derive the calibration are plotted with filled symbols.
  Solid lines (green color in electronic edition) are our linear fits used 
  to derive the transformations in Eq.~\ref{eq:acsVIcalib2}. Overplotted 
  for comparison are also transformations from Sirianni et al. (PASP, 
  submitted) derived from synthetic photometry (dotted blue and red lines) 
  and observations (dashed black lines).}
  \label{ACS_EMMI_transformations}
\end{figure}

Finally, using 67 best matches between the EMMI and ACS frames we derive the 
following transformation equations using least-squares 
fitting with an iterative
$3\sigma$ rejection algorithm:
\begin{eqnarray}
\label{eq:acsVIcalib1}
(V-F606W)& = &0.222 (\pm 0.010) \times (V-I)\cr
         &   &+ 0.072 (\pm 0.016) ~\quad\quad rms=0.046\\
(I-F814W)& =& -0.042 (\pm 0.010) \times (V-I)\cr
         &  &+ 0.130 (\pm 0.016) ~\quad\quad rms=0.050
%         &  & ~\quad\quad\quad\quad\quad\quad\quad\quad\quad rms=0.050
\label{eq:acsVIcalib2}
\end{eqnarray}
The $1\sigma$ scatter around the fit (not including points rejected 
from the fit) is shown on the right side of each equation. 
We show these fits in Fig.~\ref{ACS_EMMI_transformations} with solid lines, 
where we plot all the matched sources between ACS and EMMI images 
with error-bars; those that passed the visual inspection (67 stars), 
and were thus used in fitting are plotted with filled dots. 
The (x, y) positions
from the $F814W$ ACS image, $F606W$ and $F814W$ magnitudes in the VEGAmag system,
as well as $VI$ magnitudes from EMMI and positions for all these 
67 stars are listed in Tab.~\ref{EMMI_ACS.tab}. 

For comparison in Fig.~\ref{ACS_EMMI_transformations} 
overplotted are also 
the transformations between the ACS WFC bands and ground based VI photometry
derived by Sirianni et al.\ (submitted). Dotted lines (blue and red
respectively) are used for transformations derived from synthetic photometry 
for blue and red stars, while the dashed lines show the transformations of 
derived from the observations (see Sirianni et al. for details). 

\subsection{Completeness and Error Analysis}

\begin{figure}
\includegraphics[angle=270,width=14cm]{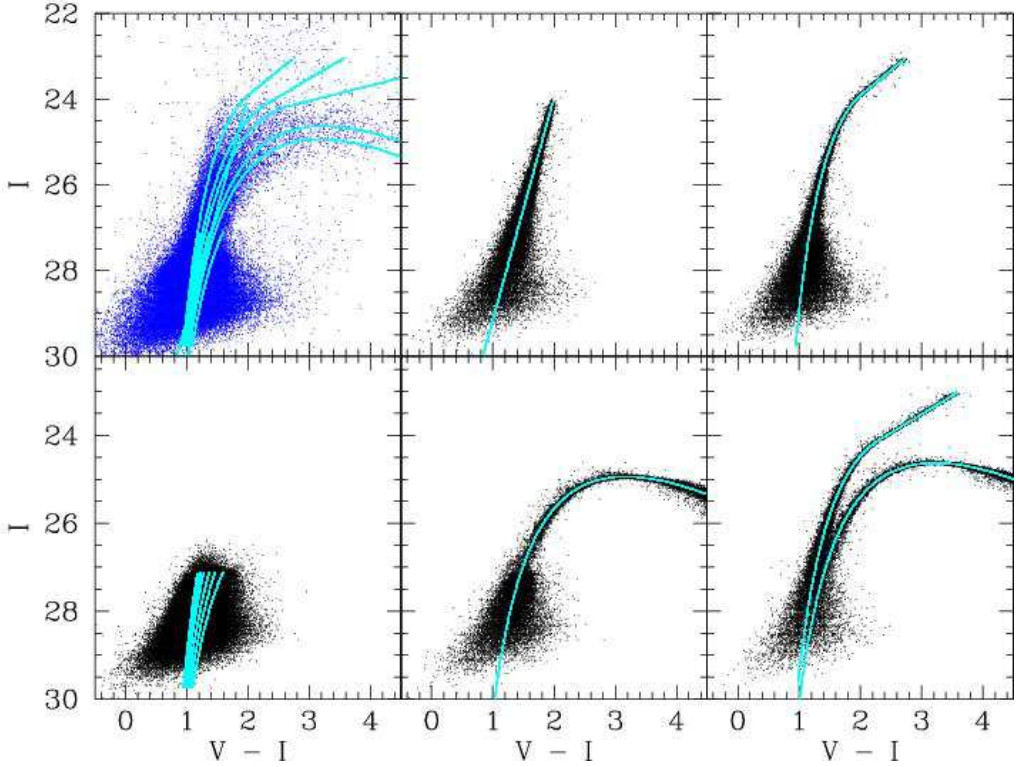}
  \caption[]{Upper left panel: observed color-magnitude diagram with simulated
	RGB sequences overplotted in light gray (cyan color in electronic 
	edition). In the other panels some of the simulated color-magnitude 
	diagrams are shown, indicating the scatter due to photometric
	errors and incompleteness as a function of magnitude and color. In 
	order to reproduce the full width of the lower RGB it was necessary to
	simulate RGB sequences with a wide range of colors corresponding to 
	$-2<[Fe/H]<0.0$ as shown in the lower left panel.}
  \label{simcmds}
\end{figure}

\begin{figure}
\includegraphics[angle=270,width=14cm]{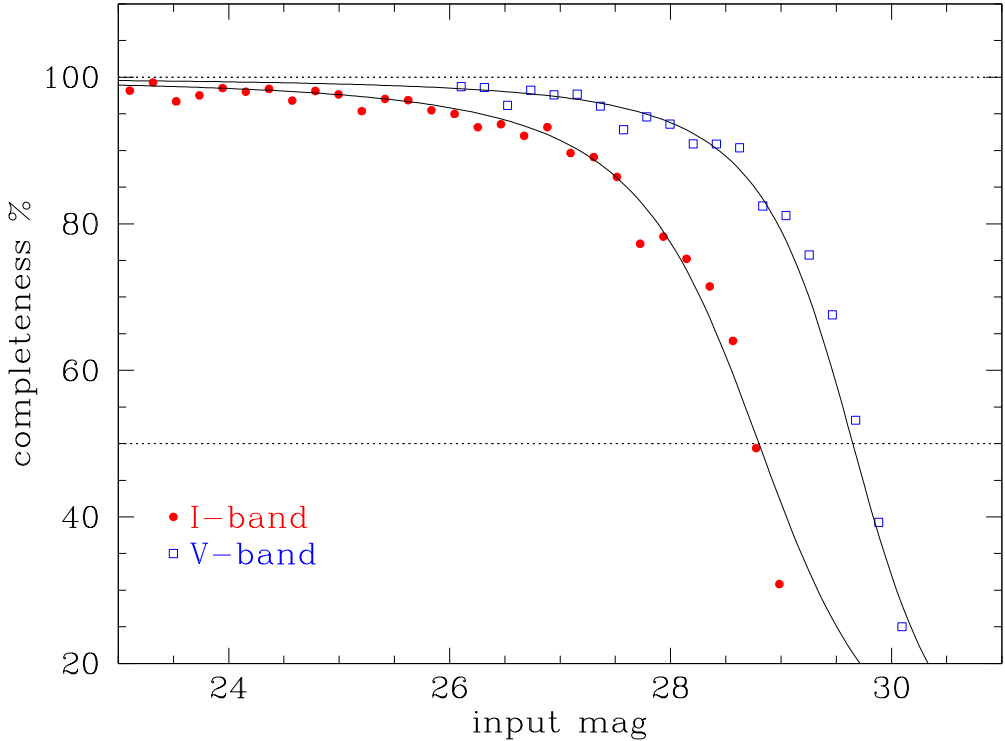}
  \caption[]{Completeness as a function of magnitude can be well fitted with 
  the analytic function for simulated stars with $V-I$ colors bluer than 
  $\sim 2.5$. This figure is a plot of completeness fraction as a function of 
  magnitude for simulations with input colors bluer than $V-I=1.5$. The fit 
  has the following parameters (see eq.~\ref{complfitequ}): $\alpha(V) = 1.1$, 
  $m_0(V)=29.65$, $\alpha(I)=0.82$ and $m_0(I)=28.80$.}
  \label{completenessfit}
\end{figure}

\begin{figure}
\includegraphics[angle=270,width=14cm]{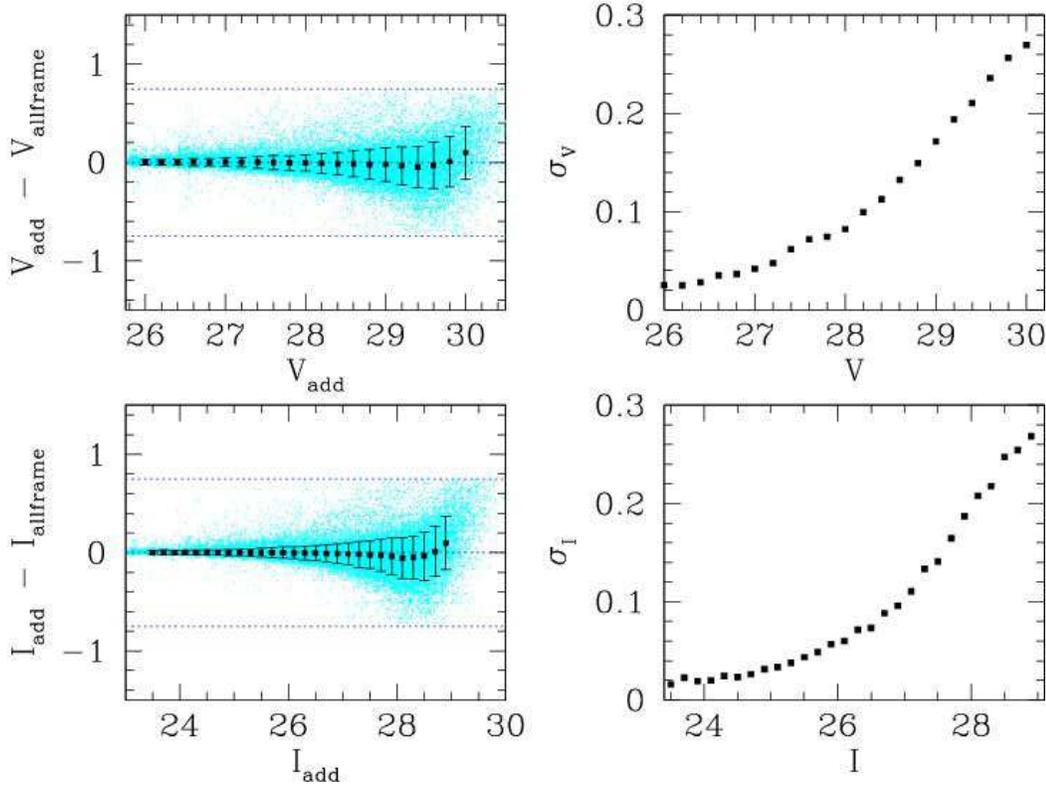}
  \caption[]{Mean error as a function of magnitude. On the left we plot
  difference between the measured and input magnitude as a function of input
  magnitude for all the stars from our completeness simulations with 
  light gray (cyan in the electronic edition) dots, while the panels on the 
  right show mean error as a function of magnitude as measured from these
  simulations. Only those simulated stars that were recovered within 1 pix 
  of the input position and that had magnitudes within 0.75 of 
  the input magnitude are considered to be recovered. There is no 
  systematic shift of the mean magnitude difference and the typical 
  $1 \sigma$ errors at the 50\% completeness magnitude limits are of 
  the order of 0.3 mag. 
	}
  \label{merror}
\end{figure}

\begin{figure}
\includegraphics[angle=0,width=15cm]{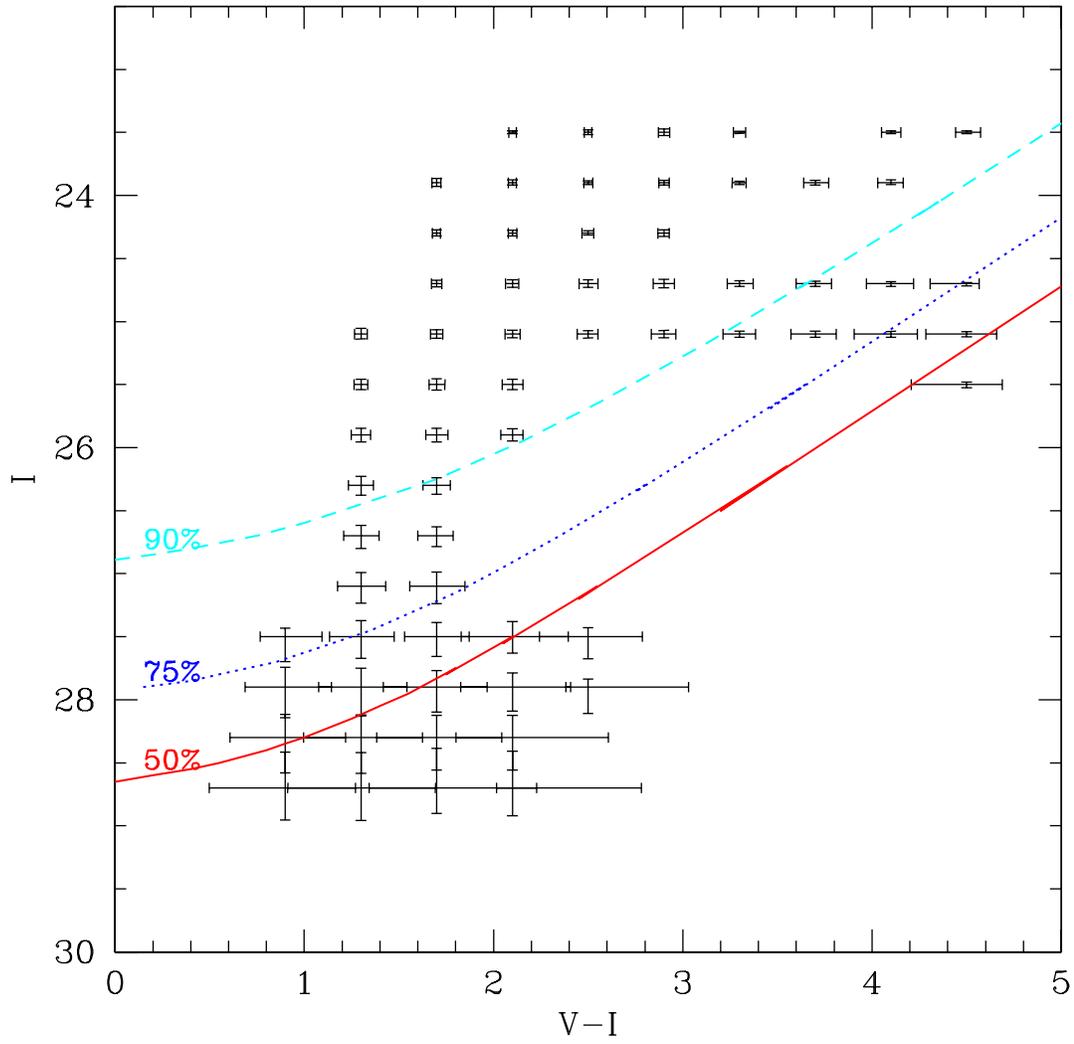}
  \caption[]{Mean error and completeness as a function of magnitude and color.
	}
  \label{cmderr.ps}
\end{figure}

The measured CMD distribution in principle suffers from both
incompleteness at the 
faint end due to loss of faint stars and an excess at the bright 
end due to blends of two or more bright stars. Completeness of our ACS 
photometry has been evaluated from normal artificial star techniques. 
Many test runs were made, in each one of which we added more than 22000 
stars to all the images distributed evenly on a grid with
minimum separation of 2.1 effective PSF radii between the added stars in order
not to increase the crowding. The starting
pixel of the grid was varied randomly in each run. 
In total, 341400 simulated stars were added in 15 experiments.
Figures~\ref{simcmds} to \ref{cmderr.ps} illustrate the results of the
completeness and error analysis.

The input magnitudes and colors of the artificial stars were chosen along 
sequences which correspond to the red giant stars in 
Galactic globular clusters spanning a range of metallicities 
$-2< \mathrm{[M/H]} < 0.0$~dex, computed using the analytical expressions 
from \citet{saviane+00}. In some cases these analytic RGB sequences were 
extrapolated to magnitudes brighter than the RGB tip magnitude in order 
to probe the completeness and photometric error of the upper part of the CMD 
where AGB stars may be expected. The input sequences are shown in light gray 
in the upper left panel of Fig.~\ref{simcmds} superimposed to the CMD 
of our ACS field (black dots). The other panels of Fig.~\ref{simcmds} 
illustrate the effect of the photometric errors and incompleteness on 
the CMD by showing both the input sequences (in light gray) and the output 
magnitudes and colors (in black). Several completeness experiments were made 
using only faint part of RGB loci as input magnitudes and colors to gain 
larger statistics in the part of the CMD where the errors and completeness 
vary most.
Clearly the wide color range of the stars in our field requires a wide 
range in metallicity. This is evident both in the upper part of
the RGB, and at the fainter magnitudes, as shown in the lower left panel.

%Examples of several completeness runs are in Figure~\ref{simcmds}. For 
%reference the true observed color-magnitude diagram is in the upper left panel, 
%overplotted with the color-magnitude sequences of the artificial stars. 
%In the lower left panel the range of artificial RGB sequences with metallicities 
%spanning $-2< \mathrm{[M/H]} < 0.0$ is plotted.  The colors of the stars 
%measured in the respective experiments reproduce the full range of observed
%colors.  

The photometric completeness is defined here as the fraction of recovered 
objects at a given magnitude $f=N_{measured}/N_{simulated}$. 
We specify that the simulated star is considered to be recovered if 
its position is measured within 1 pixel of the input position and its magnitude 
within 0.75 mag of the input magnitude (corresponding to blending of 
up to 2 stars with the same magnitude).
\citet{Fleming+95} show that the completeness 
function can be well fitted with the following analytical function:
\begin{equation}
f = \frac{1}{2} \left[ 1 - \frac{\alpha (m - m_0)}{\sqrt{1+\alpha^2(m-m_0)^2}}
\right]
\label{complfitequ}
\end{equation}
In our case, given the very wide range of colors at a given magnitude, 
the completeness at a given magnitude is also a strong function of color. 
Figure~\ref{completenessfit} shows the results of the completeness test 
limited to stars with input colors bluer than $V-I=1.5$, on which we 
superimpose the Fleming et al.\ relation with $\alpha \simeq 1.1 $
for the $V$-band and 0.8 for the $I$-band. It can be seen that, for the blue
stars, the analytic relations give a very good description of the data, 
and that the 50\% completeness is 
reached at magnitude $m_0(V)=29.65$ and $m_0(I) = 28.80$. 
Actually, these relations adequately describe the completeness for artificial
stars with colors up to 2.5.

The photometric errors, quantified as difference between the input and the
output magnitude of the artificial stars, are shown in Figure~\ref{merror}. 
The mean magnitude difference 
$\Delta mag= mag_{simulated}-mag_{measured}$ is consistent with 0 for 
magnitudes brighter than a 70\% incompleteness limit. The shift toward 
systematically brighter measured magnitudes, which signals the occurrence of
blending of the stellar images is smaller than 0.09 mag at 
the magnitude bin corresponding to 50\% incompleteness. 
The mean scatter, a measure of the 
photometric error, is smaller than 0.3 for all the magnitudes up 
to the 50\% incompleteness level as shown also on the right panels of
Figure~\ref{merror}. 

Finally, Fig.~\ref{cmderr.ps} summarizes the amplitude of the photometric errors
and the completeness levels throughout the CMD of our field. It can be noticed
that our data are fairly complete along the upper two magnitudes on the RGB over
a very wide color range. At the level of the red clump (I=28, V-I=1) our data
are still $\sim 70$\% complete.

%
%----------------------------------------------------------------------
%

\section{The Color-Magnitude Diagram}
\label{CMD}

\begin{figure}
\includegraphics[width=8cm]{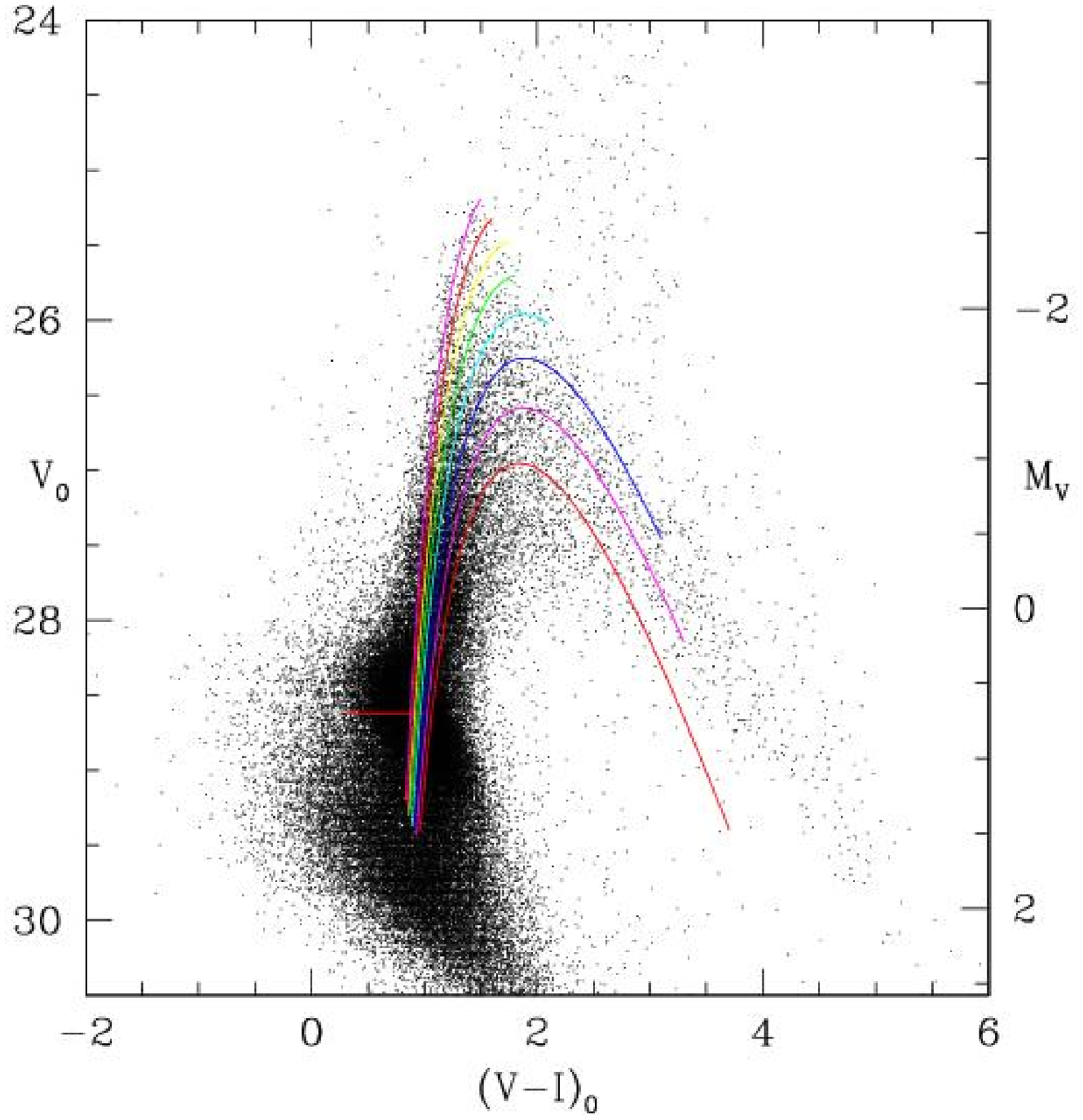}
\includegraphics[width=8cm]{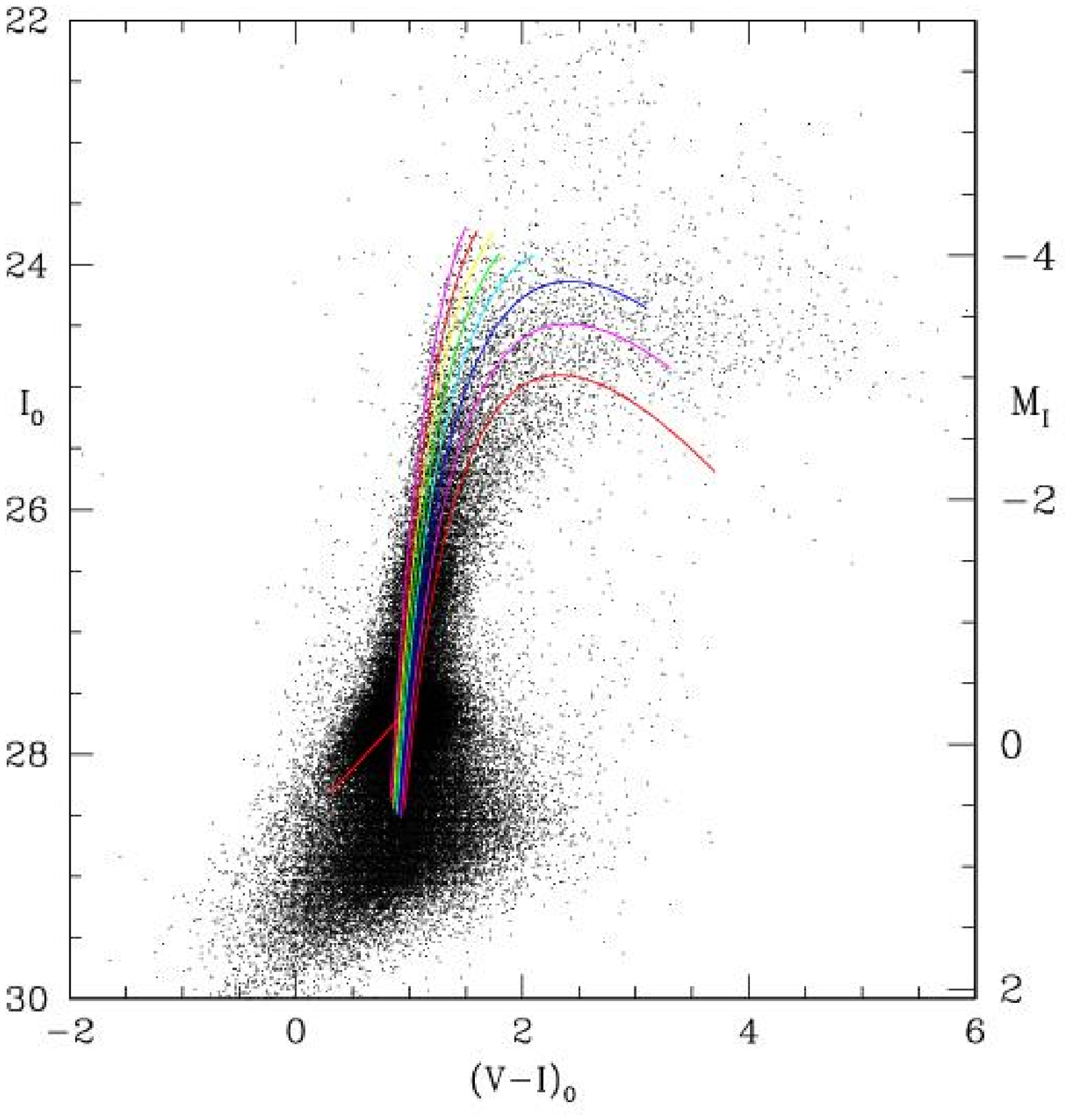}   
  \caption[]{V vs. (V-I) and I vs. (V-I) CMDs calibrated to ground based 
  system and de-reddened assuming $\mathrm{E}_{B-V} = 0.11$. 
  Overplotted are empirical fits to Galactic globular cluster RGBs ranging in
  metallicity from $-2 \leq \mathrm{[Fe/H]} \leq -0.25$ in steps of 0.25 dex 
  \citep{saviane+00}. 
  The short horizontal line indicates the position of the HB in 47 Tuc
  \citep{rosenberg+00}.
  Foreground stars have colors $\mathrm{V}-\mathrm{I} \la 3.5$ and are easily
  observed above the tip of the red giant branch stars in NGC~5128, which has 
  magnitude $\mathrm{I}=24$. Stars brighter than the tip of the RGB and redder 
  than $\mathrm{V}-\mathrm{I} \simeq 3.5$ are most probably metal-rich 
   long period variable stars in NGC~5128 halo.}
  \label{CMD_VIcalib}
\end{figure}

In Fig.~\ref{CMD_VIcalib} we plot the final CMD for our ACS field, calibrated 
to the ground-based $VI$ photometric system through the transformations above. 
In this figure as well as throughout the following discussion we have 
de-reddened the photometry assuming $E(B-V)=0.11$ \citep{schlegel} and the 
\citet{cardelli+89} reddening law.

As expected from previous studies of the NGC~5128 halo 
\citep[e.g.][]{harris+99, harris&harris00, harris&harris02}, we find that
the CMD is dominated by an old stellar population of RGB stars.
At blue colors ($V-I \la 2.5$) the RGB tip is well defined around
$I\sim 24$, but at redder colors it bends over to fainter magnitudes 
up to $I \sim 25.2$ around $V-I = 4.5$. According to our completeness
simulations we do not expect to have missed many red objects. Inspection of
our database shows that there is only one star-like object with $I<26$, and 
other two with $26<I<27$ that have not been detected on the $V$-band images.

Stars brighter than $I\sim 24$ mostly belong to the Galactic foreground 
population. 
It is easily seen that most of the Galactic stars have $0 \la (V-I)_0 \la 2.5$.
The predicted number of foreground Galactic stars in the range of 
$I=22-24$ is $\sim 5$~arcmin$^{-2}$ \citep{bahcall+soneira81}, 
yielding 57 stars in this magnitude interval in our 11.33 arcmin$^2$ 
ACS field. The Besan\c{c}on group model of the Galaxy available through the
Web\footnote{http://bison.obs-besancon.fr/modele/} 
\citep{robin+96} predicts 68 stars in this magnitude interval in
our field and a total of 415 stars with magnitudes between $20<V<29$. This is
negligible compared to the 77810 stars in our photometric
catalogue, and even around magnitude bins where the red clump and AGB bump
features are detected (see below) the Galactic contamination is of the order of
1\% at most. Above the tip of the RGB, however, 
this contamination is significant. We observe 86 stellar 
sources between $22< I_0 <24$ and $0 \leq (V-I)_0 \leq 2.5$. Most should be
foreground Galactic stars, but a few could be long-period AGB variable stars.
Finally, there are many stellar objects redder than $(V-I)>2.5$ and 
brighter than the tip of the RGB. These are most probably long period variable 
stars belonging to the metal-rich population, similar to those reported by 
\citet{rejkuba+03} in other halo fields in NGC~5128. 

At $I\sim 28$ the RGB becomes much wider and very much more heavily
populated.  We identify the excess of stars there as
the red clump (RC):  that is, core-helium burning stars, seen here for the
first time in a giant E/S0 galaxy.
In Fig.~\ref{CMD_VIcalib}, the (red) short slanted
line at $27<I < 28$ and $V-I$ from 0.4 to 1.0 indicates the expected position
of the horizontal-branch stars of 47 Tuc, the classic metal-rich 
Galactic globular cluster \citep{rosenberg+00}, shifted
to the distance modulus and reddening of NGC~5128. Notably, there is no trace 
of an extended blue horizontal branch, which would
normally be found if an old metal-poor stellar 
population were present.  However, we express a note of caution:
the large incompleteness and photometric errors 
at these faint magnitudes may prevent a weak blue horizontal
branch from being evident. At face value, the very low number of blue HB stars
indicates either a relatively young mean age or a stellar population
that has virtually no low-metallicity component.  
We argue for the latter conclusion in the discussion below.

In its general features, the CMD is strikingly similar to that of  
the lower-luminosity Local Group elliptical 
M32 \citep{grillmair+96}. The luminosity functions for these two galaxies
are compared in more detail below.

%----------------------------------------------------------------------
%

\section{The Metallicity Distribution}

\begin{figure}
\includegraphics[angle=270,width=15cm]{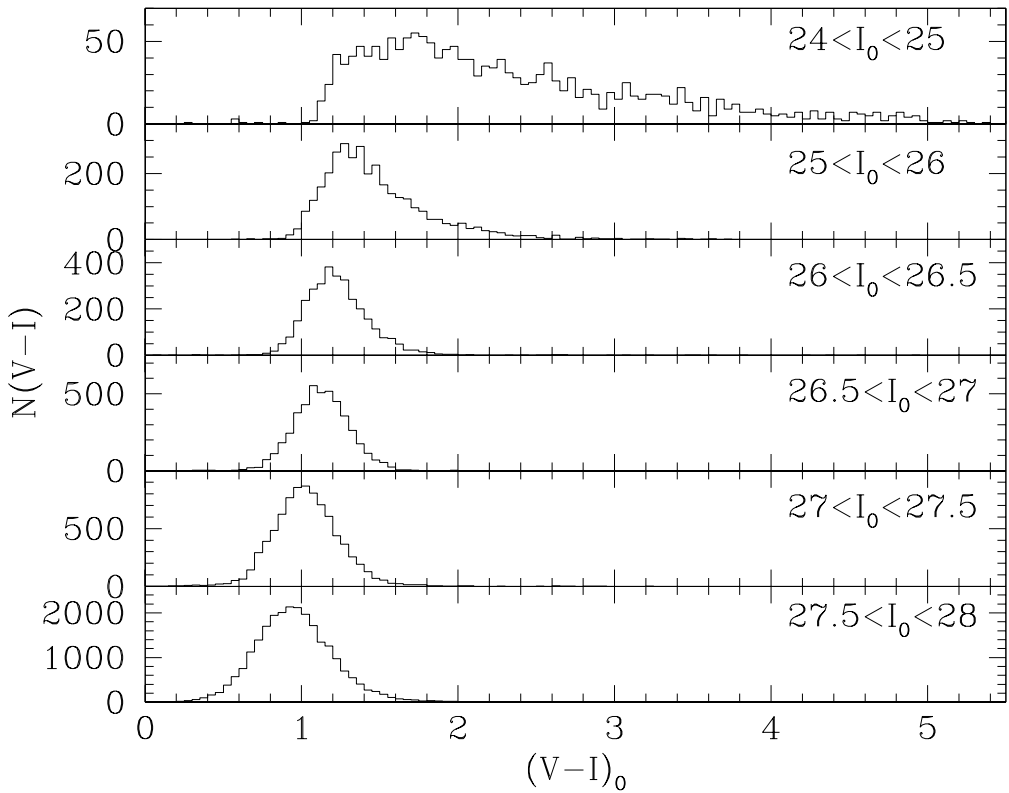}
  \caption[]{The distribution of $V-I$ color in different magnitude 
  bins along the RGB.
	}
  \label{colorlf.ps}
\end{figure}

Overplotted in Fig.~\ref{CMD_VIcalib} are empirical analytic fits to 
Galactic globular clusters RGBs with a wide range of metallicities from 
\citet{saviane+00}. Going from blue to red we plot the model RGBs from 
[M/H] = $-2$ to $-0.25$ dex in steps of $0.25$ dex. The lines trace 
rather well the metal-poor part of the RGB, but the two most metal-rich 
hyperbolas are obviously not adequate. 
It is evident that stars with almost-Solar 
metallicities are present even at this remote halo location (and it is worth
emphasizing that the {\sl minimum} true galactocentric distance of any stars in the
target field is $r_{gc} = 36$ kpc).
Also, it should be noted that due to the very red colors of the most metal-rich 
giants, in order to sample the complete metallicity distribution, 
it is necessary to reach $V$-band magnitude limits of $\sim 29$. 
The ACS data achieve this much better than the earlier WFPC2 studies.
We plot the distribution of V-I color along the RGB for  
magnitudes with %smallest errors and 
completeness larger than 70\% in Figure~\ref{colorlf.ps}.

\begin{figure}
\includegraphics[angle=0,width=15cm]{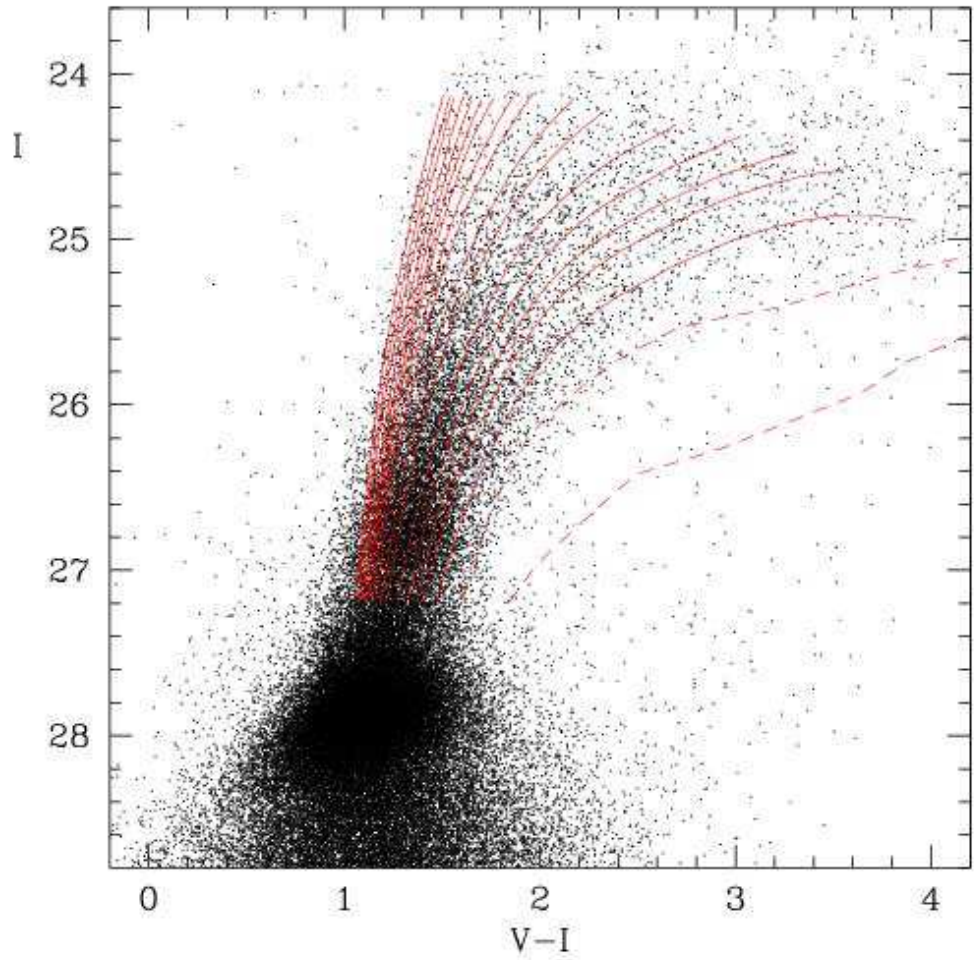}
  \caption[]{$VI$ CMD with overplotted theoretical models. The solid lines are
  1~M$_\odot$ tracks from \citet{vandenberg+00}, 
  empirically calibrated to match the RGB
  colors of Galactic globular clusters. 
  The two dashed tracks are two metal-rich tracks
  defined as described by \citet{harris&harris02}.}
  \label{cmd_fid.ps}
\end{figure}

\begin{figure}
\includegraphics[angle=0,width=15cm]{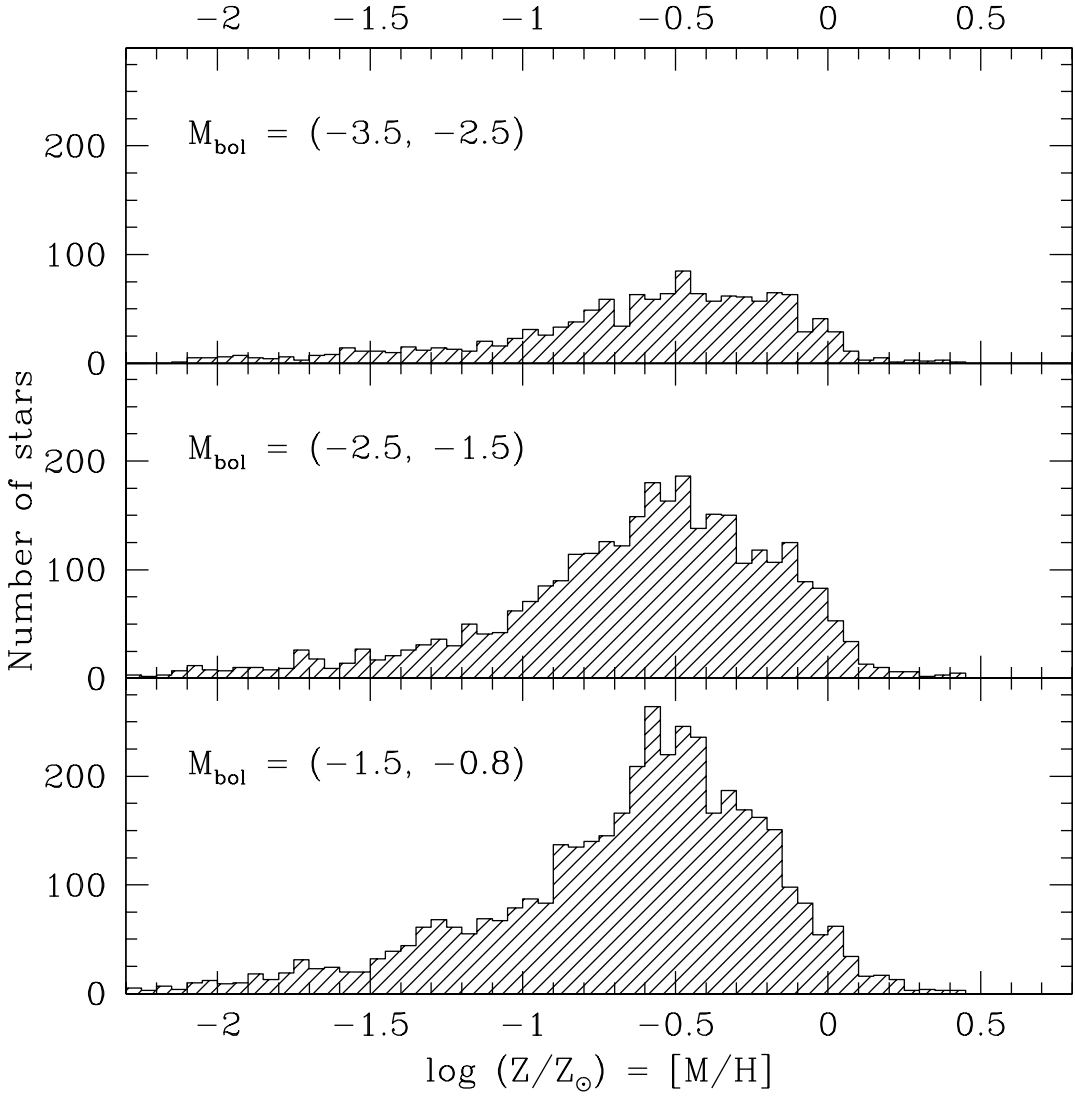}
  \caption[]{Metallicity distribution function for three different luminosity 
  bins along the RGB are shown, from the top of the RGB down to more than 
  2.5 mag fainter. Very similar spread in the three magnitude bins shows that 
  photometric errors are similar and that there is no significant 
  incompleteness up to 2.5 mag fainter than the tip of the RGB.}
  \label{feh_mbol.ps}
\end{figure}

In order to derive the MDF for this field we use the same procedure as for the
other halo fields observed with WFPC2 \citep{harris&harris00,
harris&harris02}. A finely spaced grid of red giant evolutionary tracks
\citep{vandenberg+00}, with colors calibrated using Galactic globular clusters, 
is overplotted on the CMD (Fig.~\ref{cmd_fid.ps}). 
Interpolation within the grid is done in $[M_{bol}, (V-I)_0]$ space 
\citep[see][for detailed method]{harris&harris02} to estimate the
heavy-element abundance of each star on the upper part of the RGB.
The resulting MDF, defined as number of stars per bin in log $(Z/Z_{\odot})$, is 
shown for three different magnitude bins covering the top part of the RGB
where the photometric accuracy is highest (Fig.~\ref{feh_mbol.ps}).
Encouragingly, we see no trends in the MDF with magnitude, indicating that
there are no luminosity-dependent effects in the interpolation.

\begin{figure}
\includegraphics[angle=0,width=15cm]{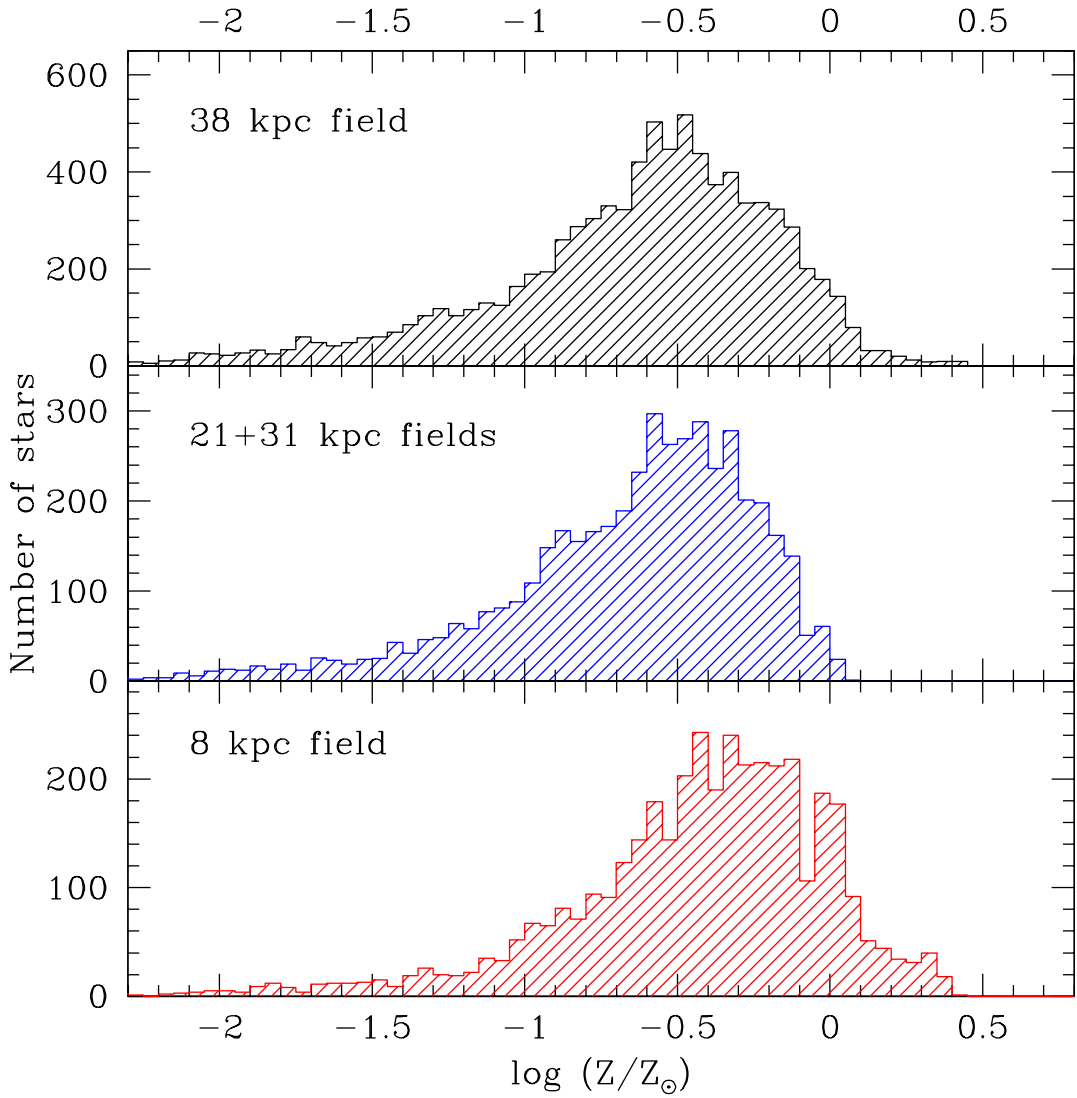}
  \caption[]{Comparison of MDF in the new ACS field (40 kpc), and in the
  	previously observed inner halo fields: middle panel shows a 
	combined MDF for 21 and 31 kpc fields \citep{harris&harris00} and the
	bottom panel is MDF for the inner (bulge) field at 8 kpc 
	\citep{harris&harris02}.}
  \label{feh_3field.ps}
\end{figure}

The combined MDF for all three bins is compared with those from the
three other halo fields previously observed with WFPC2 
\citep{harris+99, harris&harris00, harris&harris02},
in Fig.~\ref{feh_3field.ps}.  It is clear that
our outermost ACS field has an MDF very similar to those of the
mid-halo 21-kpc and 31-kpc WFPC2 fields.  As noted in \citet{harris&harris02},
the innermost 8-kpc field is significantly more metal-rich on average.
On closer inspection, we see that 
the ACS field has a more well populated high-metallicity
end than does the mid-halo MDF.  
We attribute this difference to the very large bolometric corrections for 
the reddest stars
in our CMD, which make them extremely faint in $V$ and could have
driven them below the (shallower) WFPC2 photometric limits 
\citep[see][for further discussion on this point, and attempts to derive
a completeness-corrected MDF]{harris&harris02}.
A quantitative comparison between the mean metallicities in the three fields is
presented in Table~\ref{mean_MH.tab}.
Interestingly, \citet{beasley+03} noted that their best hierarchical-merging
formation model of a giant elliptical like NGC 5128
predicts a higher fraction of most metal-rich stars than were in the
earlier CMDs, a conclusion acting in the same direction.
These comparisons show the importance of getting very deep and 
accurate photometry to 
detect the most metal-rich part of the RGB. 

Although the RGB color for ``old'' ($\sim$ 5 Gy or more) stellar populations 
is primarily sensitive to metallicity, some comments should be made about
the potential effects of age differences within the sample \citep[for a 
more detailed discussion see][]{salaris+girardi05}.
The grid of RGB tracks used to derive our MDF was normalized
to the $(V-I)$ colors and luminosities of real globular clusters in the Milky
Way \citep[see][]{harris&harris00, harris&harris02}, and thus 
implicitly assumes a uniform age for the NGC 5128 halo population of
about 12 Gy (that is, the same as the mean age for the classic Milky Way GCs).
What if the population is significantly younger?
To test the effects of a simple shift in the {\sl mean} age of the halo stars,
we have repeated the same grid interpolations using \citet{bvdb01}
isochrones of two different ages:  8 Gy and 12 Gy.  The effect of reducing
the age to 8 Gy is essentially to move our 12-Gy MDF {\sl redward} (that is,
to higher mean metallicity) by almost exactly 0.1 dex, while preserving the
overall histogram shape.  That is, for ages larger than about 8 Gy, we conclude
that the mean metallicity and internal MDF dispersion are insensitive to
age differences. An additional caveat one has to keep in mind regarding the 
derived MDF is that it contains not only RGB, but also AGB stars, which are 
on average bluer, hence skewing the MDF to a lower average value than the 
true [M/H] distribution. In a later paper, we will return to the combined 
effects of age and metallicity in determining the full distribution of the 
stars in the CMD;
and in section 7 below, we make some initial suggestions about the possible
age range within the NGC 5128 halo population based on the features of the 
CMD (the helium-burning clump and AGB bump) that are more sensitive to age.

\section{The Luminosity Function}

\begin{figure}
\includegraphics[angle=270,width=15cm]{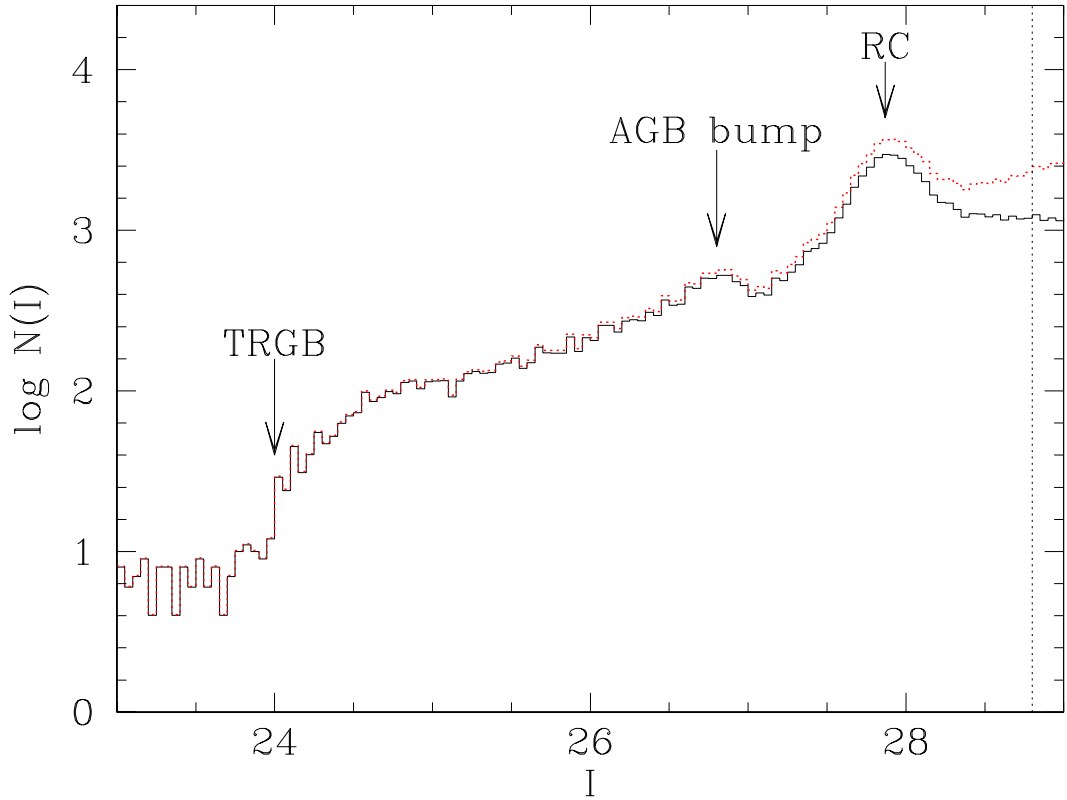}
  \caption[]{The measured $I$-band luminosity function (not corrected for 
  extinction) is plotted with (black) solid line, while the (red) dotted 
  curve shows the completeness corrected luminosity function. The vertical 
  dotted line indicates 50\% completeness magnitude of $I=28.80$. The main 
  features are indicated by arrows: tip of the RGB (TRGB) at $I_0 \sim 24.0$, 
  AGB bump at $I_0 \sim 26.77$ and red clump (RC) at $I_0 \sim 27.87$.}
  \label{loglfcomplI.ps}
\end{figure}

The $I$-band luminosity function of the observed halo field is in
Table~\ref{N5128LF.tab} and is shown in  
Figure~\ref{loglfcomplI.ps}. 
The measured $I$-band luminosity function (not corrected for extinction) is
plotted as the black solid line, while the (red) dotted curve shows the 
completeness corrected luminosity function. The vertical dotted
line indicates 50\% 
completeness magnitude of $I=28.80$. The main features are indicated by arrows:
the RGB tip, the AGB bump, and the red clump.

\subsection{Comparison with the M32 luminosity function}

\begin{figure}
\includegraphics[angle=270,width=15cm]{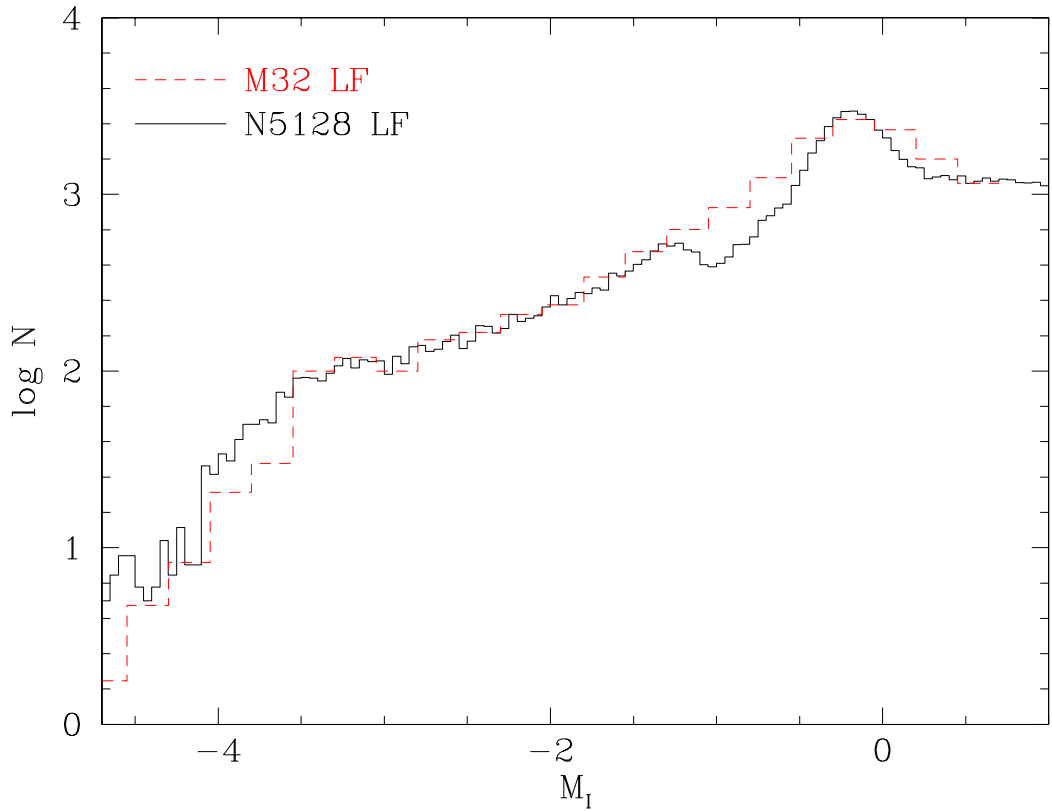}
  \caption[]{The comparison between the LFs of NGC~5128 derived here 
  (solid line) and M32 LF from \citet{grillmair+96} (dashed line).}
  \label{lflog_m32.ps}
\end{figure}

The nearest example of an elliptical galaxy is M32. 
Spectroscopically it is similar to faint ellipticals, 
but its very high surface brightness and proximity to M31 make
it somewhat of an unusual case and a very difficult object to study. 
\citet{grillmair+96} presented its $VI$ CMD obtained with WFPC2 on HST, 
which appears very similar to the CMD of our field in NGC~5128. 
The comparison of the luminosity function in NGC~5128 and in 
M32 is presented in Figure~\ref{lflog_m32.ps}, where the latter  
was scaled using the
distance modulus of $(m-M)^{M32}=24.43$ and reddening of 
$E_{B-V}^{M32}=0.08$. The scaling of the number counts was chosen arbitrarily. 

The RGB tip and the RC features have very similar luminosities in the two 
galaxies, implying similar ages if the metallicity distributions are 
the same. However, according to \citet{grillmair+96}  the M32 population 
may be more metal-rich than the halo population in NGC~5128. 
It should be noted though that the observations of M32 sampled the inner
$1'-2'$, which may well have higher metallicity than its outer regions. 
The absence of the AGB bump feature in the M32 luminosity function 
is most probably due to larger photometric errors and larger magnitude 
bins along the luminosity function. 

\subsection{RGB tip}
\label{sect:TRGB}

The RGB tip, the feature due to the helium flash luminosity, 
is often used as a 
``standard candle'' to determine galaxy distances. Its brightness in the 
I-band is (almost) independent of age and metallicity for metal-poor
($\mathrm{[Fe/H]}\leq -0.7$~dex), hence blue RGB stars \citep{DA90,harris+99}. 
Selecting only blue stars in our CMD ($(V-I)<1.8$) we use a Sobel 
filter \citep[see][]{lee+93} to measure the RGB tip magnitude. The filter is 
applied to luminosity functions computed with various bin sizes, ranging 
from 0.015 to 0.025 with a step of 0.005, each yielding one measurement of 
the RGB tip magnitude. The average of all these measurements is 
$I(TRGB) = 24.05 \pm 0.05$. This is in agreement with the estimate of
$24.1 \pm 0.1$ from the 21-kpc field \citep{harris+99} and 
$24.0 \pm 0.1$ as measured by \citet{soria+96} for a 10-kpc field.

\begin{figure}
\includegraphics[angle=0,width=15cm]{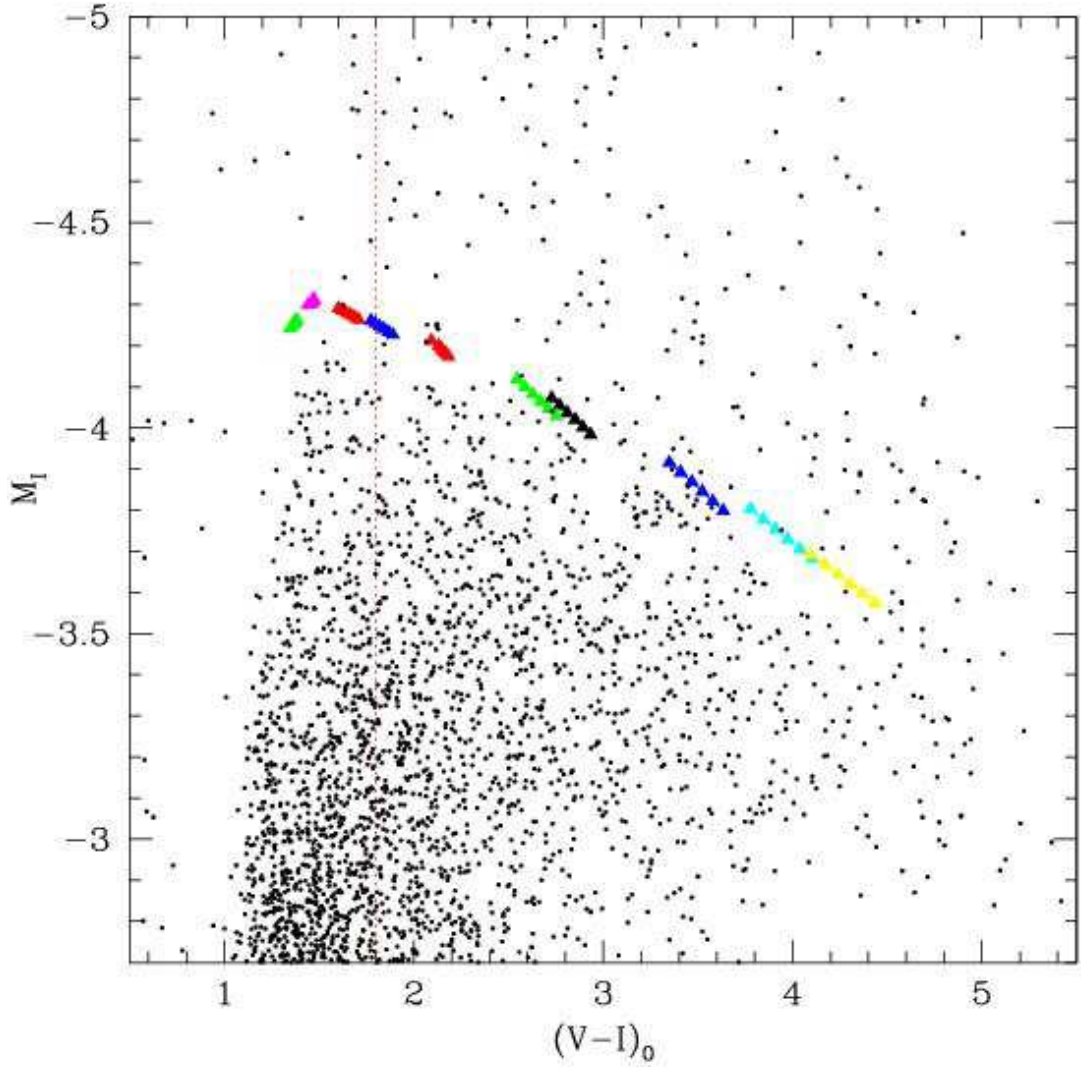}
  \caption[]{The comparison between the observed $VI$ CMD and the RGB tip
  magnitude as a function of age and metallicity. The RGB tip magnitude
  predictions from \citet{pietrinferni+04} stellar evolutionary models are
  plotted as (colored) triangles for 10 different metallicities 
  (Z=0.0001, 0.0003, 0.001, 0.002, 
  0.004, 0.008, 0.01, 0.0198, 0.03 and 0.04) and ages ranging from 7 to 
  12 Gy. Models of different ages for a fixed metallicity are 
  connected with lines (see electronic edition for the 
  color version of this figure).
	}
  \label{obs_teo_RGBtipCMD.ps}
\end{figure}

Our data provide an excellent opportunity to check the dependence of the 
RGB tip magnitude on metallicity in the models. 
With respect to globular clusters, this
dataset has the following advantages: (1) very good sampling of the RGB tip
(where evolutionary lifetimes are short); (2) wide range of metallicity; and (3)
same distance. The obvious disadvantage is the fact that the stars at RGB 
tip in our field have a range of ages and metallicities and the models 
predict some dependence of RGB tip I-band magnitude on star-formation history
and in particular on the age-metallicity relation \citep{salaris+girardi05}. 
The comparison of the model predictions for the
RGB tip magnitudes in the range of ages between $7 \leq age \leq 12$~Gy and
for 10 different heavy metal abundances (Z=0.0001, 0.0003, 0.001, 0.002, 
0.004, 0.008, 0.01, 0.0198, 0.03 and 0.04), computed using stellar evolutionary
tracks of \citet{pietrinferni+04}, is shown in 
Figure~\ref{obs_teo_RGBtipCMD.ps}. The triangles of different colors connected
with lines (see the electronic version for the color figure) 
show RGB tip magnitudes for models of fixed metallicity and ages
between 7 and 12 Gy. The RGB tip $I$-band magnitude is fainter for 
older ages at a fixed metallicity. The vertical dotted line is drawn 
at $V-I=1.8$, the reddest color for the measurement of the
RGB tip magnitude. These models 
reproduce well the slope of I magnitude which gets fainter as V-I gets redder
or the stars more metal-rich. However, they predict a brighter RGB tip 
magnitude than observed by $\sim 0.2$ mag.
We believe it is largely due to the adopted $\mathrm{BC}_I$ scale
\citep[see detailed discussion in][]{salaris+girardi05}.
Part of the difference may also come from
the combined uncertainty of the RGB tip magnitude
measurement ($I^{TRGB}$) from the data and the uncertainty in 
the adopted distance modulus of NGC~5128 
\citep{harris+99, rejkuba04}, which is of the order of 0.15~mag.

\subsection{AGB bump}

The AGB bump feature is related to the evolutionary behavior of
stellar models when helium is exhausted in the core and ignited in a
shell. As the central helium burning approaches the end, the core
contracts rapidly, until a helium burning shell is ignited.  The
ensuing violent expansion pushes the star temporarily out of thermal
equilibrium and the model climbs rapidly to brighter luminosities
\citep{renzini77}.
When the helium shell is fully ignited, thermal equilibrium is
restored in the stellar envelope and evolution proceeds on a nuclear
timescale. This marks the base of the slow evolution on the AGB, or
the AGB bump. Due to the very short lifetime of the AGB phase, 
the AGB bump feature can be detected only when dealing with large 
samples of stars. From the observational point of view it was 
discussed by \citet{gallart98} and \citet{ferraro+99}.
\citet{alves+sarajedini99} and \citet{cassisi+01} have discussed 
theoretical predictions for AGB bump magnitude as a function of age 
and metallicity of the parent population.

\begin{figure}
\includegraphics[angle=270,width=14cm]{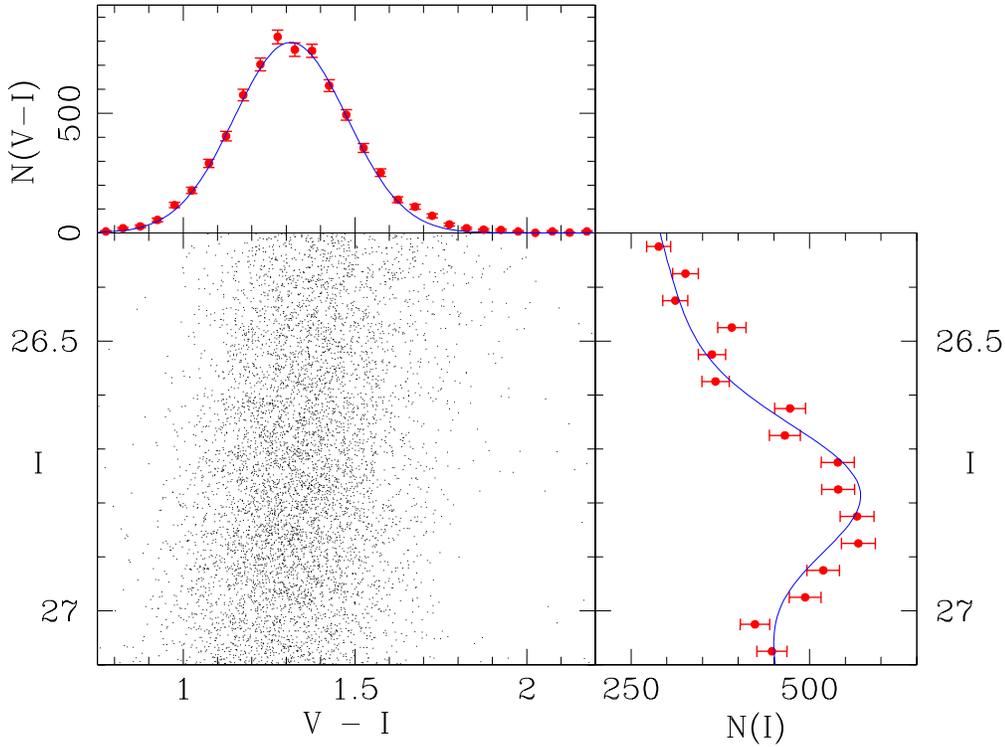}
  \caption[]{A zoom on the CMD around the AGB bump feature. On the right 
  the I-band luminosity function around the AGB bump is fitted with a 
  Gaussian plus a straight line. The peak is at $I=26.77 \pm 0.01$ and 
  $\sigma = 0.12$. Above we show a Gaussian fit to the color distribution 
  around the AGB bump.}
  \label{AGBbumpI.ps}
\end{figure}

The bump in the $I$-band luminosity function in  
Figure~\ref{loglfcomplI.ps} around $I \sim 26.8$ occurs at the right magnitude
expected for an AGB bump. At this level our photometry is $\sim 90$\% 
complete.
A linear relation plus a Gaussian provides a good fit to
the I-band luminosity function in the range $26<I<27.1$ 
(Fig.~\ref{AGBbumpI.ps}).
The best-fit Gaussian has the following parameters: 
mean magnitude, corresponding to the observed magnitude
of the AGB bump, is $I = 26.77 \pm 0.01$, corresponding to $M_I^{AGBb}=-1.31$, 
and the standard deviation is 0.12 mag.
The mean color of the AGB bump is $V-I=1.313 \pm 0.003$ with a width of 
the Gaussian fit of 0.166 mag. The mean magnitude and color errors in this
region of the CMD are of the order of 0.085 mag. 

The appreciable color range
present in the AGB clump smears out this feature in the V-band luminosity 
function. Still, a small bump is observed also in $V$ and the best fitting 
Gaussian plus a straight line give $V=27.97 \pm 0.07$,
corresponding to $M_V^{AGBb}=-0.29$, and $\sigma = 0.13$. 
The somewhat wider AGB bump in $V$ is the 
consequence of the higher sensitivity on metallicity of this
feature in the $V$ band, rather 
than in the $I$ band (see below) and of the wide metallicity
distribution of stars contributing to the feature. It is not obvious how much 
a possible age spread in the halo also widens the bump. 
A more complete analysis will be done 
later through  simulations of the entire CMD. Here we compare the 
mean magnitude and color of the AGB bump and the red clump (see next 
section) to stellar models in order to determine a plausible mean 
age of the halo stars.

\subsection{Red clump}

\begin{figure}
\includegraphics[angle=270,width=14cm]{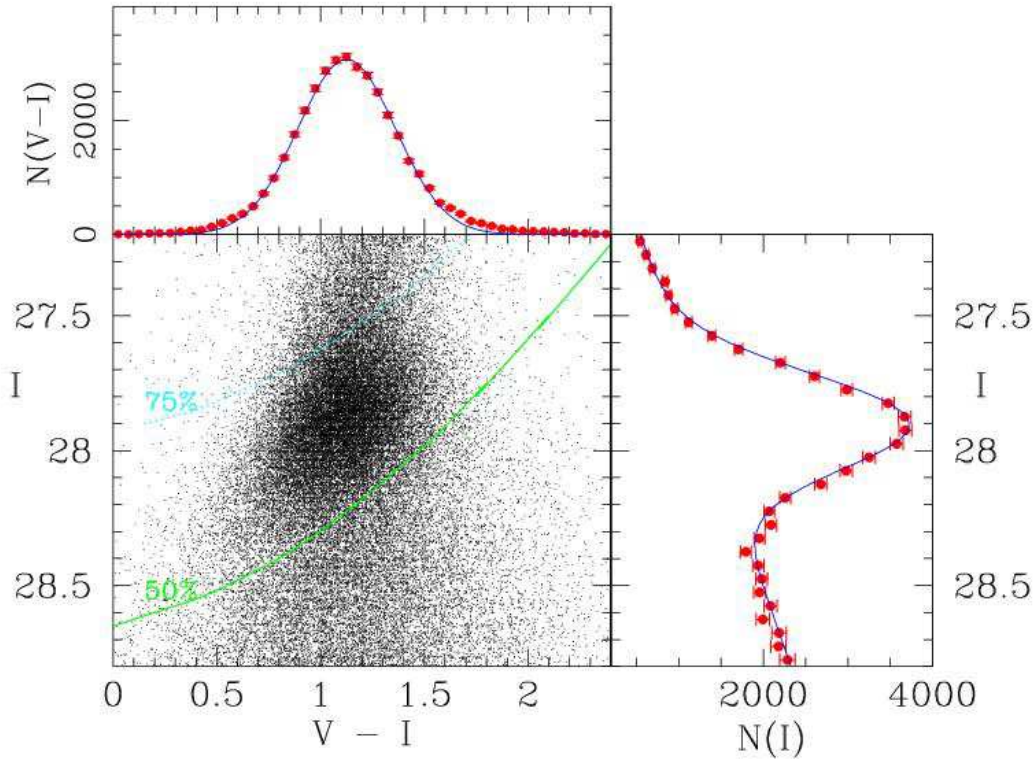}
  \caption[]{A zoom on the CMD around the red clump feature. On the right the 
  I-band luminosity function around the RC is fitted with a Gaussian plus a 
  straight line. The peak is at $I=27.873 \pm 0.002$ and $\sigma = 0.18$.
  Above we show a Gaussian fit to the color distribution.}
  \label{RCI.ps}
\end{figure}

From the MDF, and from the absence of defined blue horizontal branch, 
it can be concluded that the metal-poor component is virtually
absent in this galaxy, even in the outermost halo. On contrary, in M31
halo, which has been shown to have significant intermediate-age as well as
metal-rich component \citep{durrell+04,brown+03}, blue horizontal branch is 
well defined and extends $\sim 1$ mag bluer than the average red giant branch 
color. In NGC~5128, most of the core helium 
burning stars are located in the red clump (RC) attached to the RGB sequence. 
A Gaussian plus a straight line again provides a good fit to the I-band 
luminosity function here as well, in the range 
$27.2<I<29$ (Fig.~\ref{RCI.ps}). 
The best fit Gaussian has a mean magnitude, 
corresponding to the observed magnitude of the RC, of $I(RC) = 27.873 \pm 0.002$ 
(corresponding to $M_I^{RC}=-0.21$) 
with $\sigma = 0.18$. The mean color of the RC is $V-I=1.126 \pm 0.001$
with a spread of $0.23$. The mean magnitude and color errors in this 
region of the CMD are 0.18 and 0.2 mag. Thus a large part of the width of the 
RC is due to photometric errors and not only due to a mix of populations. The
completeness level of the photometry varies across the RC magnitudes between 50\% and
75\% (Fig.~\ref{RCI.ps}).

In the $V$-band the luminosity function of the RC 
can be fitted with a Gaussian at 
$V=28.951 \pm 0.003$ and with $\sigma=0.20$. This corresponds to 
$M_V^{RC}=0.69$~mag.

\section{Comparison with models}

In this section we compare the magnitudes of the AGB bump and the RC features
with the most recent stellar evolutionary tracks from \citet{pietrinferni+04},
in order to derive the mean age for the halo stars in our field. 
It is obvious from the upper RGB, as well as from the previous work,
that there is a wide range of metallicities present in the field; but
some age spread cannot be excluded \citep{rejkuba+03, peng+04b}.
Therefore we need to map the theoretical location of the AGB bump and the Red
Clump for populations in a range of age and metallicity. To do that we 
construct theoretical luminosity functions of Simple Stellar
Populations (i.e. assembly of stars with the same age and metallicity, SSP)
as follows. 

For an SSP, the number of stars in any magnitude bin (j) along the post main
sequence (MS) portion of its representative isochrone can be approximated 
fairly well by:
\begin{equation}
\Delta N_j^{SSP} = \phi(M_{TO}) |\dot{M}_{TO}| \Delta t_j
\label{eq_deltanj1} 
\end{equation}
where $M_{TO}$ is the turn-off mass of the isochrone, $\dot{M}_{TO}$ is the 
rate of change of $M_{TO}$ for increasing age of the isochrone, 
$\phi(M_{TO})$ is the initial mass function (IMF) evaluated
at the turn-off mass, and $\Delta t_j$ is the evolutionary
lifetime spent by $M_{TO}$ within the j-th magnitude bin \citep{renzini81}.
This equation follows from approximating the post MS section of the isochrone
with the evolutionary track with mass equal to $M_{TO}$.

Adopting, as usual, $\phi(M) \propto M^{-\alpha}$, Eq. \ref{eq_deltanj1}
can be written as \citep{greggio02}:
\begin{equation}
\Delta N_j^{SSP} =  M^{SSP} \beta_{\alpha} \Delta t_j
\label{eq_deltanj2} 
\end{equation}
where $M^{SSP}$ is the total stellar mass of the SSP and 
$\beta_{\alpha}$ is the  evolutionary flux at the turn-off per unit mass
(i.e.\ number of stars leaving the MS per
unit time in a SSP of 1 $M_\odot$):
\begin{equation}
\beta_{\alpha} = f_\alpha M_{TO}^{-\alpha} |\dot{M}_{TO}| .
\label{eq_deltanj3} 
\end{equation}
In this equation $f_\alpha$ is a factor which depends 
only on the IMF, while both 
$\beta_{\alpha}$ and $\Delta t_j$ depend on age and metallicity.

Using this formalism it is possible to construct theoretical luminosity
functions for the post MS phases of SSPs from a set of stellar tracks.
We have considered \citet{pietrinferni+04} evolutionary models for
masses $\leq 2.5~M_\odot$ and the
available heavy element abundances (i.e.\ Z=0.0001, 0.0003, 0.001, 0.002, 
0.004, 0.008, 0.01, 0.0198, 0.03 and 0.04).
These models were computed using a scaled solar
distribution for the heavy elements and the solar model has 
$Z=0.0198$ and $Y=0.273$ \citep[see][for details]{pietrinferni+04}.
The turn-off ages of these models range between 
$0.4 \la \tau \la 17$~Gy.
For each chemical composition we have constructed the relation between 
turn-off mass and age ($\tau$) which turned out to be
well represented by a parabola on the $\log M - \log \tau$ plane\footnote{We
actually fit the relation between stellar mass and age at the base of the
RGB which is more appropriate to derive the luminosity function 
along the evolutionary
phases relevant to this application.}. We have thus obtained the analytic
$\beta_{\alpha}$ functions for the various chemical compositions. 
The luminosity function of an
SSP of unitary mass is then computed as 
$\Delta n_{j}^{SSP} = \beta_{\alpha} \Delta t_j$,
in which the quantities $\Delta t_j$ are read from the track whose
evolutionary lifetime is equal to the age of the SSP.

\begin{figure}
\includegraphics[angle=270,width=15cm]{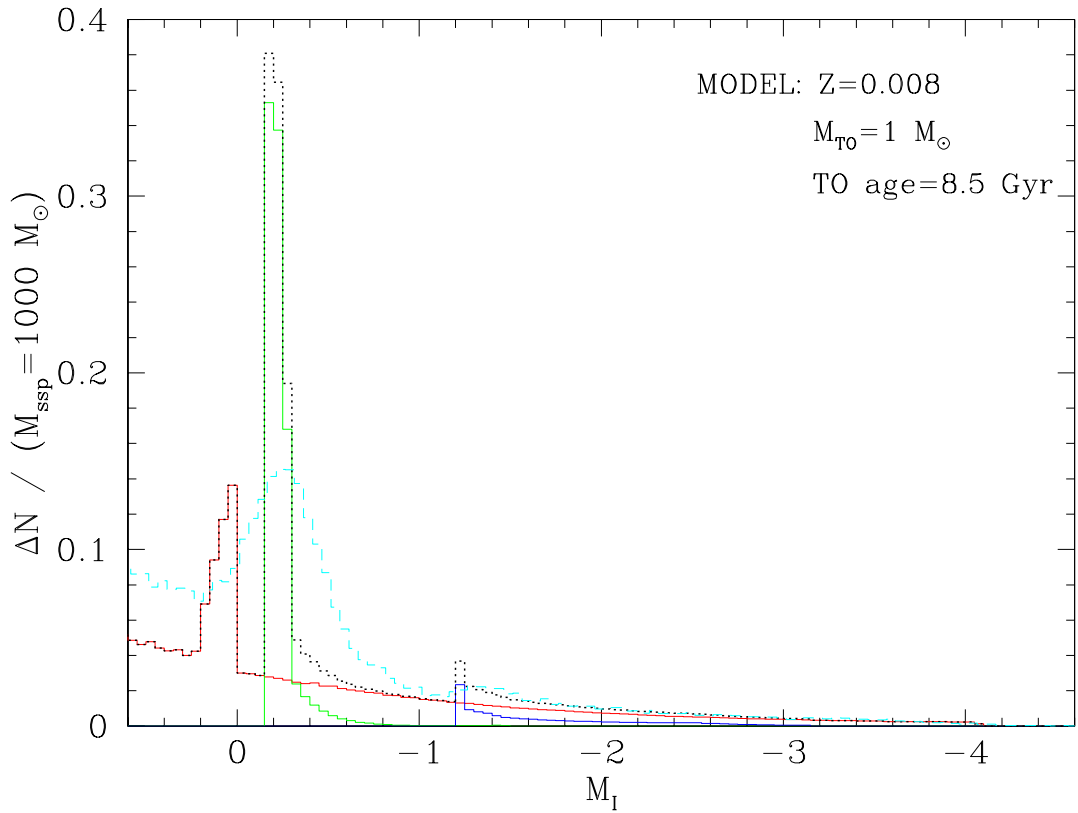}
  \caption[]{Comparison between the observed luminosity function 
  (dashed cyan line) and the luminosity function predicted for an 
  age of 8.5 Gy and heavy element abundance of $Z=0.008$, 
  computed with the evolutionary tracks of \citet{pietrinferni+04}. 
  We plot separately the RGB (red), core helium burning 
  (green) and AGB luminosity functions (blue). The sum of these three 
  components is plotted with the dotted (black) line. 
  This theoretical LF predicts  $M_I(AGBb) \simeq -1.2$, 
  $M_I(RC) \simeq -0.8$, and $M_I(RGBbump) \simeq 0.1$~mag. From
  this comparison it is apparent that the RC feature has most probably also a
  contribution from RGB bump (RGBbump) stars.
  (see electronic edition for the color version of this figure)}
  \label{obs_teoLF.ps}
\end{figure}

Figure~\ref{obs_teoLF.ps} shows one such example of a theoretical 
luminosity function for an age of 8.5 Gy and heavy element abundance of
$Z=0.008$. We plot separately the RGB (red), helium burning (green), and AGB
portions (blue), as well as the total luminosity function 
(black dotted line). The (cyan) dashed
line shows our observed, completeness corrected luminosity function, 
scaled between
$-2<M_I<-2.5$ to match the theoretical luminosity function (in this particular
case the adopted scaling factor is 25321). 
The width of the observed Red Clump feature is partly due to 
photometric errors, and partly reflects the age and metallicity spread.
In addition, the comparison between the data and the models indicates that
the RC has most probably also a contribution from the RGB bump feature.
However, since the expected number of stars in the RGB bump is relatively 
small compared to that in the RC, we neglect its effect on the magnitude
location of the Red Clump.

\begin{figure}
\includegraphics[angle=270,width=11cm]{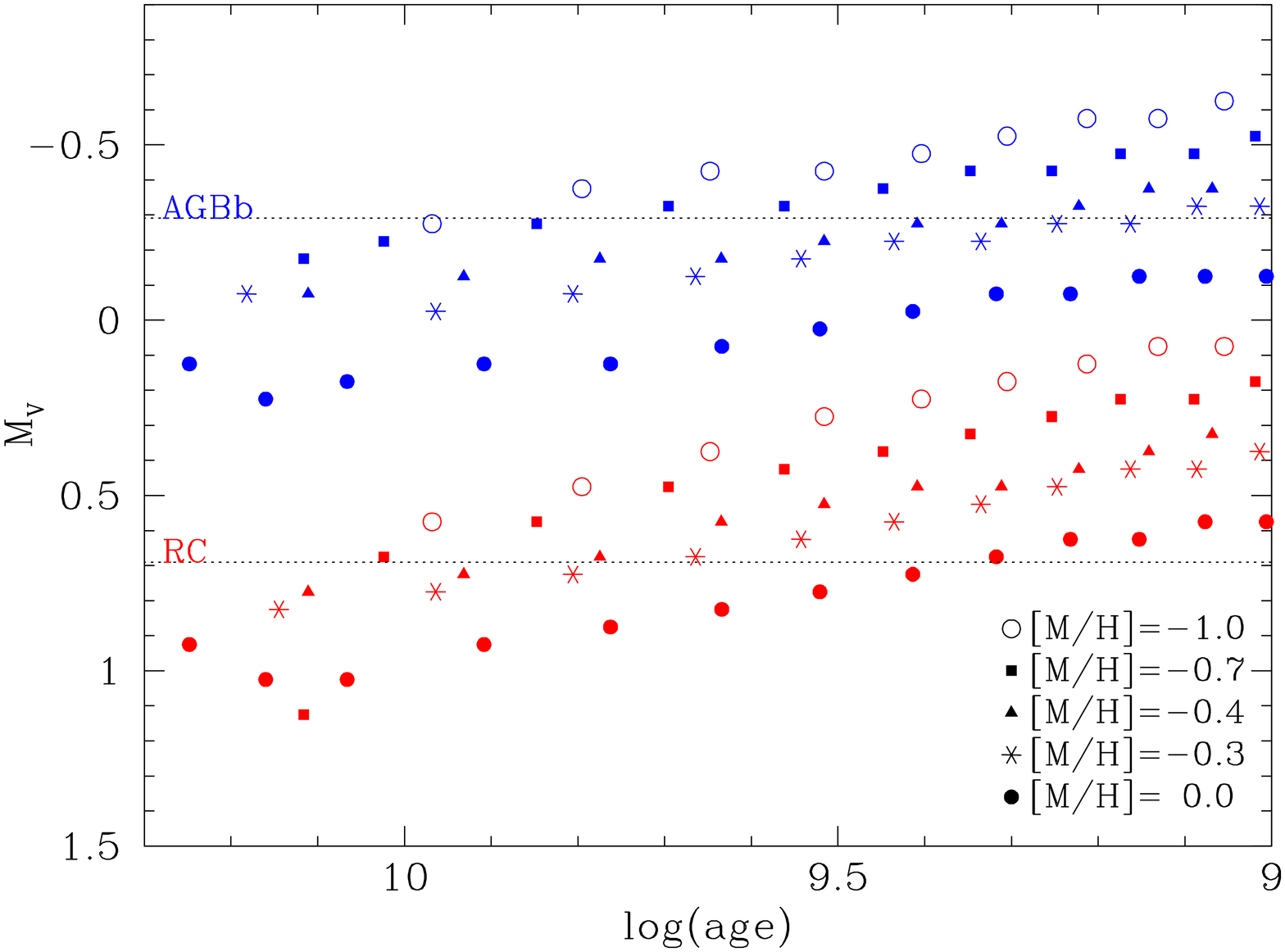}\\
\includegraphics[angle=270,width=11cm]{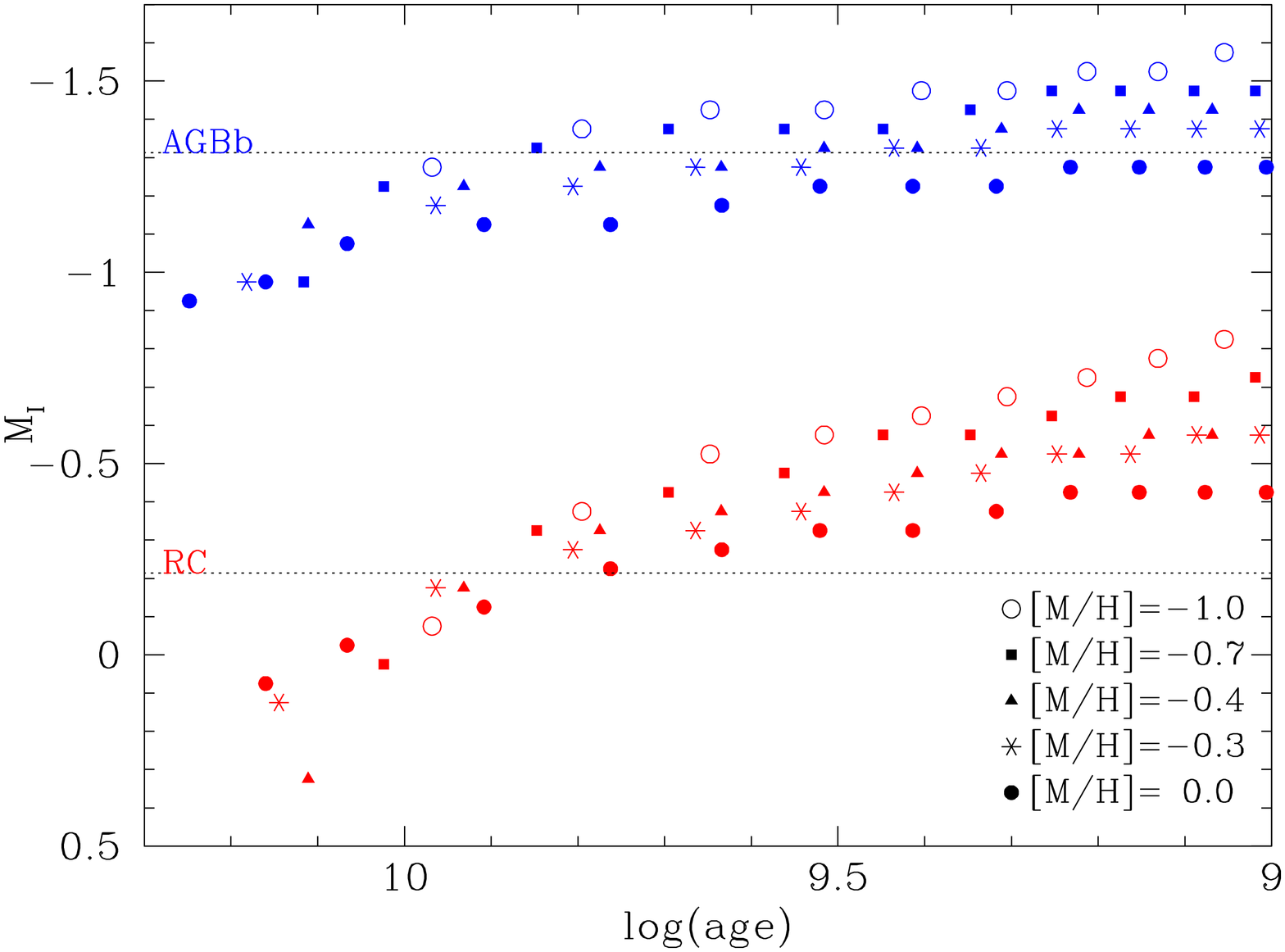}
  \caption[]{$V$ (top) and $I$ (bottom) magnitudes of AGB bump
  and red clump features predicted by the models are plotted as a function of
  age for 5 metallicities, covering most of the range of observed 
  MDF in the halo of NGC~5128. For comparison the observed AGB bump and RC
  magnitudes are indicated with horizontal dotted lines. 
	}
  \label{fig:bumpsVI}
\end{figure}

Figure~\ref{fig:bumpsVI} shows the comparison between the measured $V$ (top
panel) and $I$ magnitudes (bottom panel) of AGB bump and RC features in
NGC~5128 halo with models of five different metallicities plotted as a
function of age. The model $M_I$ and $M_V$ values are the peak magnitudes
obtained from the theoretical luminosity functions described above.
The $V$ band has a greater sensitivity to {\it metallicity}, but
the {\it age} sensitivity is similar for both bands. It is thus useful to compare the
values coming from observations in both bands. However, it should
be kept in mind that the AGB bump magnitude has much larger uncertainty, 
in particular in the $V$ band. 
We can derive a first estimate of the 
mean age of the stellar population in our field by considering that
its average metallicity is $\mathrm{[M/H]}=-0.6$ dex, which is 
close to that of
the models represented as filled squares. The location 
of both features in both bands indicates an average age around 8 Gy.

\begin{figure}
\includegraphics[angle=270,width=11cm]{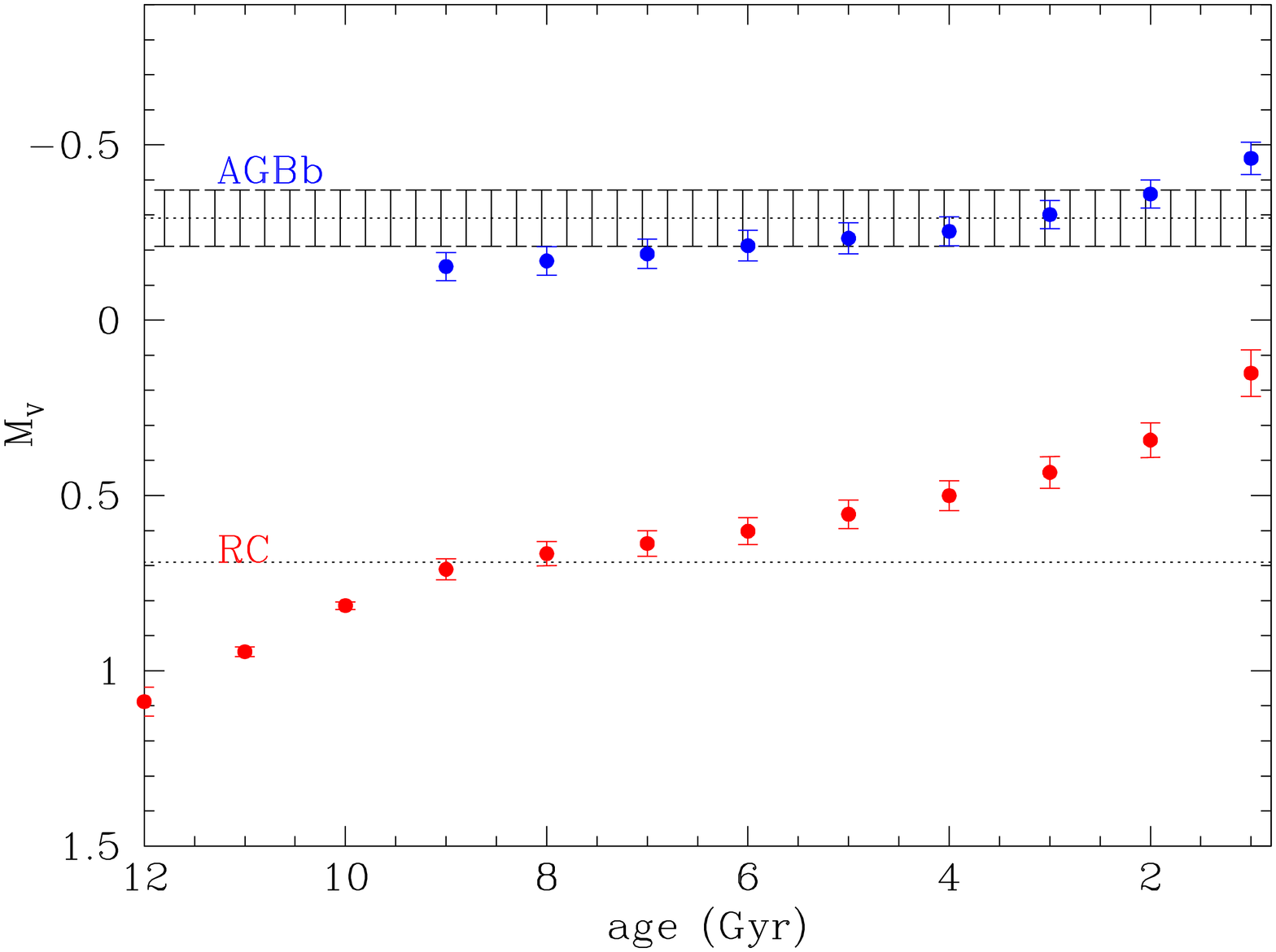}\\
\includegraphics[angle=270,width=11cm]{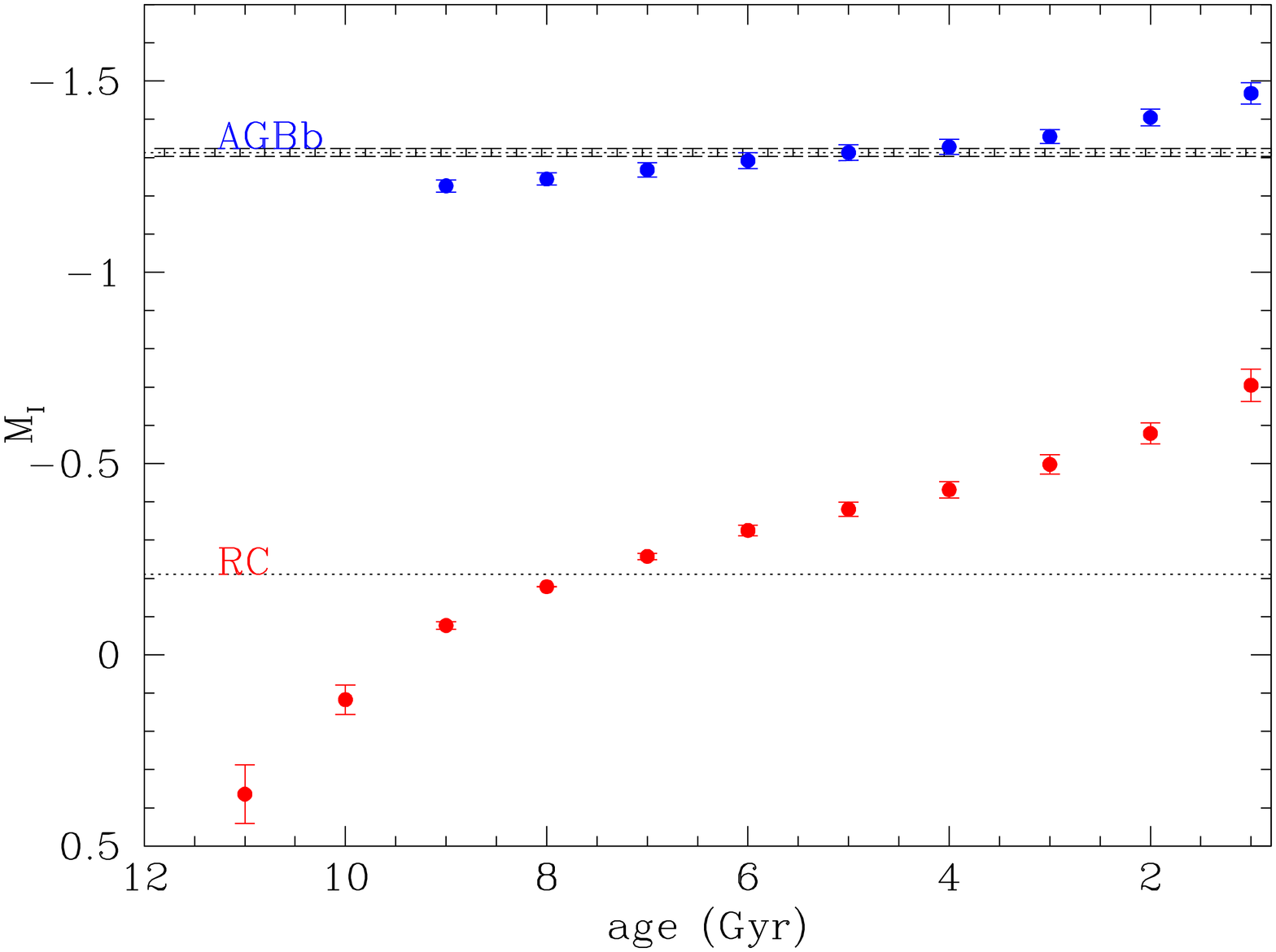}
  \caption[]{Average $V$ (top) and $I$ (bottom) magnitudes of AGB bump
  and red clump features predicted by the models, obtained by convolving the
  metallicity dependency with the measured metallicity distribution of the 
  RGB halo stars (see Eq.~\ref{eq:convolve}),  are plotted as a function of
  age. The measured AGB bump (AGBb) and RC magnitudes are shown as dotted 
  horizontal lines. We also plot $1\sigma$ error-bars for the uncertainty 
  in the AGB bump mean magnitude fit. The error-bars on the model points 
  are calculated from assumed $\pm 0.1$~dex uncertainty in the measured 
  metallicity distribution.}
  \label{fig:bumpsVI_interpZ}
\end{figure}

A better estimate of the average age of the stars in our field can be
obtained by computing the average magnitude of the RC and of the AGB
bump expected for the metallicity distribution determined from our
data:
\begin{equation}
\langle Mag(age)\rangle = \int_{\Delta Z} N(Z) Mag(Z, age) dZ
\label{eq:convolve}
\end{equation}
This estimate will be adequate insofar the age spread is small, and as
long as our derived metallicity distribution does not depend strongly
on age. As we argued earlier this is the case for the 
ages of 8~Gy and older. Figure~\ref{fig:bumpsVI_interpZ} shows the
comparison between the observed AGB bump and RC magnitude, and the results
of the application of Eq.~\ref{eq:convolve} at various ages. 
The $1\sigma$ uncertainty in the 
measured AGB bump magnitude is indicated with shaded stripes, while the
measurement of the mean RC magnitude has negligible errors due to much larger
number of stars and is thus shown as a horizontal line. It should be kept in
mind that this is only the statistical error-bar for the measurement of 
the mean RC and AGB bump magnitude in our data, but it does not include 
systematic errors due to reddening and distance uncertainties, 
which are of the order of $\pm 0.15$ mag. 
The error-bars of the model data points are calculated 
by using the Eq.~\ref{eq:convolve} with $\pm 0.1$~dex shift in the 
observed metallicity distribution and they do not include any uncertainty 
due to the adopted bolometric corrections or uncertainties in the 
models themselves, 
which are particularly relevant for the AGB phase \citep{cassisi+01}. 

For metal-poor populations with $\mathrm{age} > 9$~Gy, the AGB bump feature is 
not visible in the theoretical luminosity functions, 
and there is only a very weak feature present 
for more metal-rich and old stars. 
Thus we do not plot the model predictions for mean magnitudes of the
AGB bump for older ages.

The $V$ band magnitude of the RC favors an average 
age for the halo stars of 8.5~Gy, while the $I$ band measurement 
agrees better with an average age of $7.6$~Gy. As shown above the 
$V$ band is more sensitive to metallicity than the $I$-band. Thus even 
small errors in the metallicity distribution (due to inclusion of AGB 
giants, or due to interpolation with tracks that have different age than 
the true distribution) can have large effect on the $V$-band models. 
This can be
also appreciated by the size of the error-bars on the model points. 
The AGB bump magnitudes in both bands seem to favor slightly younger ages, but
the shallower dependence of the AGBb with age and its larger internal
uncertainty do not exclude the average age of $8^{+2}_{-2.5}$~Gy, where the
error-bars include $0.15$~mag uncertainty in the reddening and 
distance modulus.

\begin{figure}
\includegraphics[angle=270,width=11cm]{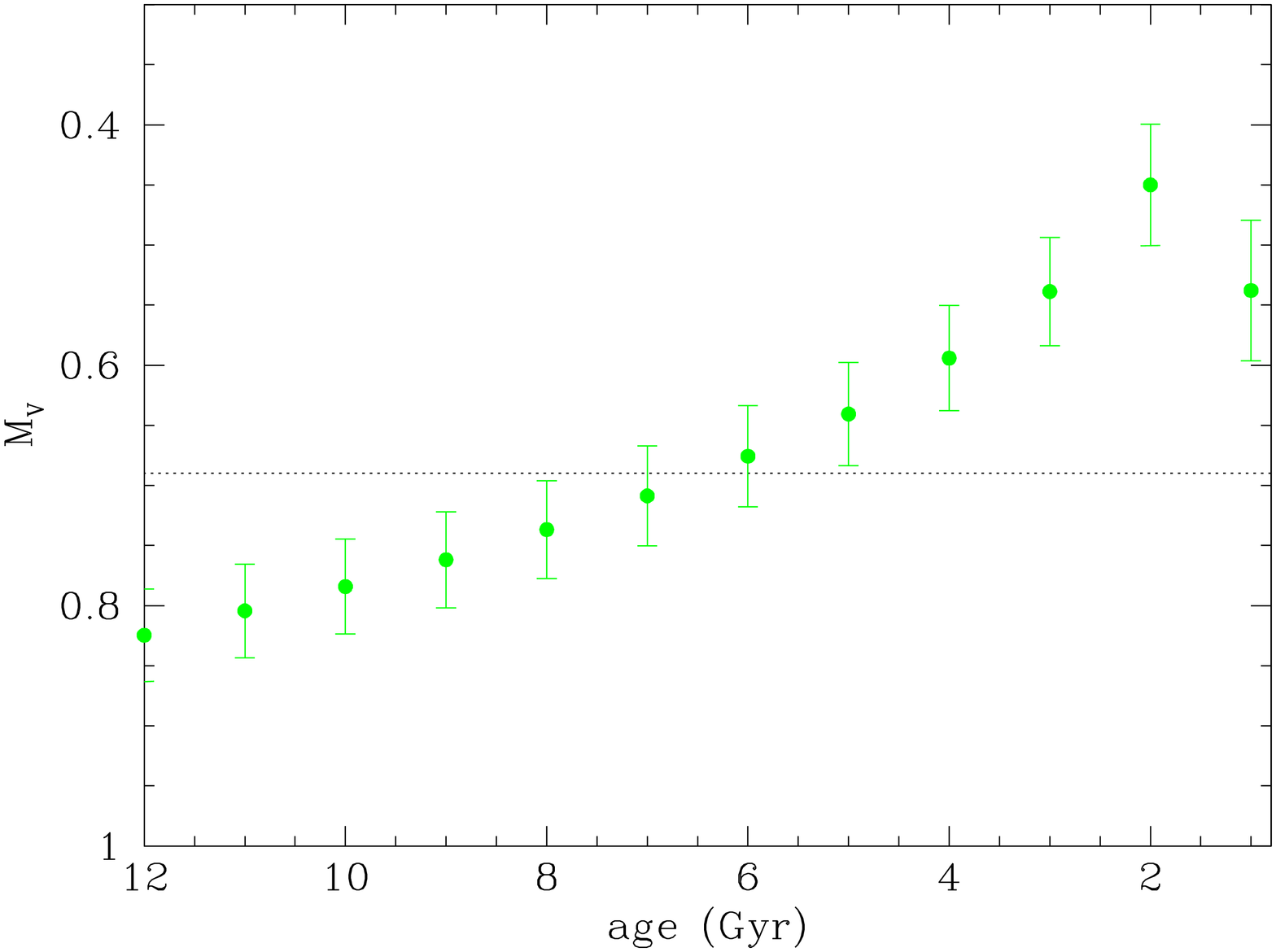}\\
\includegraphics[angle=270,width=11cm]{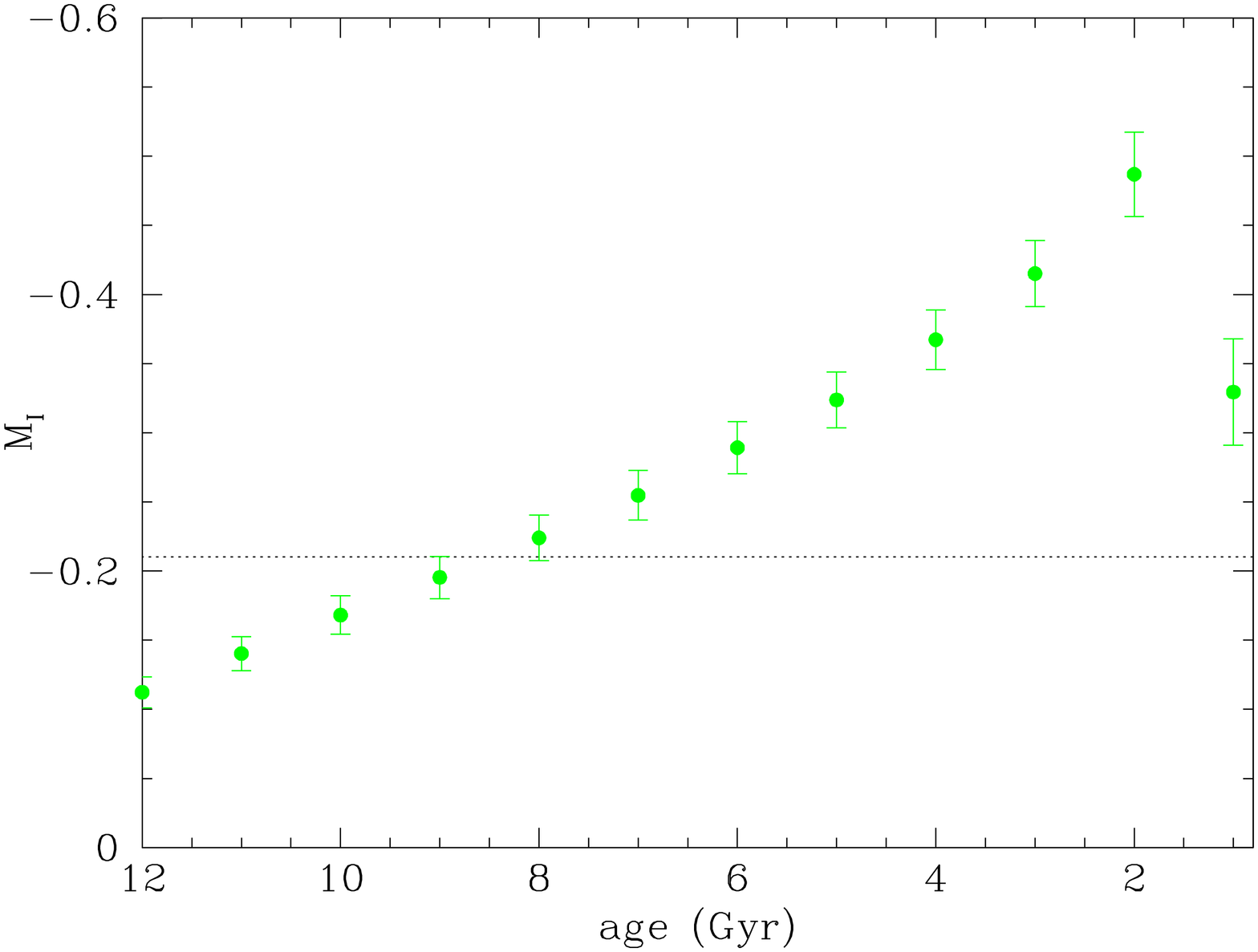}
  \caption[]{Average $V$ (top) and $I$ (bottom) magnitudes of 
  red clump predicted by the models of \citet{girardi+00,girardi+salaris01}, 
  obtained by convolving the
  metallicity dependency with the measured metallicity distribution of the 
  RGB halo stars (see Eq.~\ref{eq:convolve}),  are plotted as a function of
  age. The measured RC magnitudes are shown as dotted 
  horizontal lines. The error-bars on the model points 
  are calculated from assumed $\pm 0.1$~dex uncertainty in the measured 
  metallicity distribution.}
  \label{fig:bumpsVI_interpZ_G00}
\end{figure}

We note that even though there is a rather large discrepancy between
the observed tip of the RGB and that predicted by the models of 
\citet{pietrinferni+04}, this is most probably largely due to uncertain
bolometric corrections for these cool giants (see sect.~\ref{sect:TRGB}). 
The uncertainty in bolometric corrections for AGB bump and RC evolutionary 
phases is expected to be much smaller.
In order to test the uncertainty due to adoption of the particular set 
of models we also compare the observed magnitude of the RC in our field with
another set of model predictions, using the results published by 
\citet{girardi+salaris01}, which are based on Padova group models
\citep{girardi+00}. The average metallicity weighted RC magnitude as a 
function of age (points in Fig.~\ref{fig:bumpsVI_interpZ_G00})
was calculated again using the Eq.~\ref{eq:convolve}. 
Figure~\ref{fig:bumpsVI_interpZ_G00} shows the result. As before, the error 
bars on the model data points show the effect of  0.1 dex shift in the 
observed MDF and the dotted horizontal line indicates the measured magnitude of
the red clump. Contrary to the models of \citet{pietrinferni+04}, the $V$-band
magnitude of the RC indicates younger average age of 6.5~Gy, while the $I$-band
magnitude would imply 8.5~Gy for the average age of the halo stars. The
difference in age predicted in the two photometric bands for Padova models
\citep{girardi+00} 
is almost twice as large as that for Teramo models \citep{pietrinferni+04}, 
and we add 1~Gy error due to that. However, this does not 
change our conclusion about the average age of $8^{+3}_{-3.5}$~Gy
for the halo stars in NGC~5128.

%flat relation of average magnitude vs. age and much larger errors in the
%measurements do not exclude the average age of $ 8^{+2}_{-2.5}$~Gy, 
%predicted by the RC magnitudes, with the error-bars that take into 
%account the uncertainty in the reddening and distance modulus.

%                                                                      
%----------------------------------------------------------------------
%
\section{Conclusions and summary}
\label{discussion}

We present a very deep CMD of a remote halo field in NGC~5128, 
located 38~kpc south from the center of the galaxy. The high resolution 
and sensitivity of the ACS WFC instrument at HST allowed us for the first 
time to detect the AGB bump and RC features in a  E/S0 galaxy. 
From the colors of RGB stars we derive the metallicity distribution, 
assuming no age dependency (and an old age) of the models.

The metallicity distribution is broad and moderately rich with average
$\mathrm{[M/H]}=-0.64$~dex and a $1\sigma$ spread of $0.49$~dex. 
It is very similar to the MDFs derived from earlier observations of 
21 and 31~kpc fields, except for a slightly larger number of most 
metal-rich stars, which were missed in the previous observations.  
This metallicity distribution is convolved with the models
for AGB bump and RC magnitudes which are dependent both on age and 
metallicity to derive an average age of 
$\sim 8^{+3}_{-3.5}$~Gy for the halo stars.

In the models with ages $\ga 10$~Gy there are very few AGB bump stars
compared to the RC or horizontal branch. In metal-poor 
($\mathrm{[M/H]}<-1$~dex) and old ($\mathrm{age}>9$~Gy) models the 
AGB bump is not present at all, although
\citep{ferraro+99} report the observations of an AGB bump for 4 Galactic 
globular clusters with $\mathrm{[M/H]}<-1$. Thus an
alternate interpretation could be that a part of the
population is ``old'' ($\ga 10$ Gy) and that the
AGB bump feature appears due to a second major star-burst event that 
happened $\sim 5-7$~Gy ago. The presence of these younger stars would 
brighten the average magnitude of the RC \citep{salaris+03}, 
which would thus appear as a 
feature with a mean age around 8~Gy. 
This would be in agreement with some previous studies of 
NGC~5128 halo stars which suggested a 10\% 
of intermediate-age stars \citep{soria+96,marleau+00,rejkuba+03}.
%, and
%would provide the stellar population corresponding to some spectroscopically 
%confirmed intermediate-age globular clusters \citep{peng+04b}.
In a following paper we will investigate how much spread in age might be
present in the halo stars through 
the simulations of the complete CMD. This will also allow us to derive a 
more accurate and internally consistent 
MDF and to account for the effect of photometric errors 
as well as the presence of some probable 
RGB bump contribution to the average magnitude of the RC. 

\acknowledgments
We would like to thank Rolly Bedin for helpful discussions about the 
ACS photometry calibration and for sharing the results of 
his work before the publication. 
WEH and GLHH acknowledge financial support through research grants
from the Natural Sciences and Engineering Research Council of Canada.
EWP acknowledges support from NSF grant AST 00-98566.  Support for this
program was received from NASA through grant GO-9373.06-A from the Space
Telescope Science Institute, which is operated by the Association of
Universities for Research in Astronomy, Inc., under NASA contract NAS
5-26555.

\clearpage

\begin{table}
\caption[]{List of all the matched sources in ACS and EMMI images that passed visual inspection and were 
used to derive 
calibration Equations~\ref{eq:acsVIcalib1} and \ref{eq:acsVIcalib2}.}
\centerline{
\begin{tabular}{|rrrrrr|rrrrrr|}
\hline
\multicolumn{6}{|c|}{ACS} & \multicolumn{6}{c|}{EMMI}\\
\multicolumn{1}{|c}{x}&  \multicolumn{1}{c}{y} & \multicolumn{1}{c}{F606W}&\multicolumn{1}{c}{$\sigma_{F606W}$} &
\multicolumn{1}{c}{F814W}&\multicolumn{1}{c}{$\sigma_{F814W}$} &  \multicolumn{1}{|c}{x}&  \multicolumn{1}{c}{y}& 
\multicolumn{1}{c}{V}&\multicolumn{1}{c}{$\sigma_{V}$} & \multicolumn{1}{c}{I}&\multicolumn{1}{c|}{$\sigma_{I}$} \\
\multicolumn{1}{|c}{pix}& \multicolumn{1}{c}{pix}& \multicolumn{1}{c}{mag}&\multicolumn{1}{c}{mag}&
\multicolumn{1}{c}{mag}&\multicolumn{1}{c}{mag}&\multicolumn{1}{|c}{pix}& \multicolumn{1}{c}{pix}&
\multicolumn{1}{c}{mag}&\multicolumn{1}{c}{mag}&\multicolumn{1}{c}{mag}&\multicolumn{1}{c|}{mag} \\
\hline		   
   3937.97 &	154.85 &  21.330 &  0.017 &  20.404 &  0.022 &    328.96 &    397.22 &  21.590 &  0.012 &   1.312 &  0.030   \\ 
   2164.61 &	192.27 &  21.508 &  0.009 &  19.435 &  0.015 &    292.14 &    659.77 &  22.085 &  0.018 &   2.570 &  0.021   \\ 
   3615.89 &	279.24 &  20.772 &  0.012 &  19.806 &  0.014 &    339.63 &    447.76 &  21.152 &  0.011 &   1.215 &  0.017   \\ 
   3719.73 &	402.83 &  21.804 &  0.012 &  20.417 &  0.014 &    360.29 &    435.40 &  22.268 &  0.018 &   1.770 &  0.027   \\ 
   1572.44 &	410.62 &  23.330 &  0.010 &  21.142 &  0.010 &    310.11 &    752.41 &  23.967 &  0.076 &   2.732 &  0.086   \\ 
   2776.76 &	420.25 &  22.374 &  0.006 &  20.443 &  0.011 &    340.30 &    574.91 &  22.968 &  0.025 &   2.446 &  0.032   \\ 
   3218.67 &	448.54 &  21.487 &  0.012 &  19.497 &  0.018 &    355.10 &    510.42 &  22.043 &  0.018 &   2.515 &  0.021   \\ 
   3155.36 &	638.67 &  21.917 &  0.014 &  19.854 &  0.019 &    381.67 &    524.34 &  22.434 &  0.025 &   2.586 &  0.028   \\ 
   3283.54 &	720.19 &  21.309 &  0.010 &  20.485 &  0.014 &    396.71 &    507.44 &  21.618 &  0.009 &   1.042 &  0.020   \\ 
   3996.31 &	796.90 &  22.468 &  0.018 &  21.098 &  0.021 &    425.07 &    404.14 &  22.882 &  0.028 &   1.820 &  0.040   \\ 
     57.28 &	812.82 &  21.609 &  0.018 &  20.335 &  0.019 &    333.09 &    985.33 &  22.055 &  0.016 &   1.652 &  0.023   \\ 
   3267.80 &	852.76 &  22.679 &  0.010 &  21.798 &  0.011 &    415.80 &    513.06 &  23.000 &  0.034 &   1.051 &  0.070   \\ 
    325.51 &	950.30 &  21.622 &  0.009 &  20.640 &  0.011 &    359.85 &    949.15 &  22.058 &  0.018 &   1.263 &  0.027   \\ 
    345.39 &   1028.98 &  21.201 &  0.008 &  19.729 &  0.010 &    371.94 &    948.10 &  21.763 &  0.015 &   1.970 &  0.019   \\ 
    191.73 &   1060.88 &  23.073 &  0.019 &  21.415 &  0.017 &    372.84 &    971.47 &  23.640 &  0.071 &   2.204 &  0.085   \\ 
   1542.83 &   1067.58 &  22.706 &  0.017 &  21.575 &  0.016 &    406.35 &    772.59 &  23.151 &  0.041 &   1.522 &  0.058   \\ 
   1335.47 &   1218.15 &  21.641 &  0.021 &  19.999 &  0.020 &    423.49 &    806.74 &  22.135 &  0.013 &   2.151 &  0.019   \\ 
    233.39 &   1392.28 &  21.862 &  0.013 &  20.944 &  0.015 &    422.82 &    973.32 &  22.187 &  0.016 &   1.172 &  0.031   \\ 
    702.78 &   1682.18 &  21.689 &  0.009 &  20.389 &  0.010 &    476.78 &    911.16 &  22.215 &  0.022 &   1.743 &  0.031   \\ 
    734.77 &   1813.78 &  20.804 &  0.009 &  19.835 &  0.010 &    496.92 &    909.57 &  21.152 &  0.009 &   1.176 &  0.016   \\ 
    296.15 &   1833.29 &  22.180 &  0.012 &  19.935 &  0.011 &    489.35 &    974.68 &  22.831 &  0.028 &   2.881 &  0.032   \\ 
   1093.76 &   1998.74 &  23.232 &  0.007 &  21.427 &  0.009 &    532.81 &    861.25 &  23.819 &  0.069 &   2.260 &  0.087   \\ 
   3077.80 &   2333.91 &  21.175 &  0.010 &  19.471 &  0.021 &    629.62 &    577.00 &  21.756 &  0.013 &   2.275 &  0.016   \\ 
    137.91 &   2434.48 &  22.074 &  0.018 &  21.470 &  0.015 &    574.10 &   1012.42 &  22.079 &  0.029 &   0.968 &  0.065   \\ 
   2877.56 &   2481.52 &  21.177 &  0.016 &  19.376 &  0.025 &    646.58 &    610.11 &  21.790 &  0.013 &   2.458 &  0.015   \\ 
   2415.01 &   2489.81 &  21.095 &  0.014 &  20.436 &  0.019 &    636.80 &    678.41 &  21.391 &  0.010 &   0.861 &  0.019   \\ 
   1053.66 &   2514.91 &  22.620 &  0.012 &  22.000 &  0.011 &    607.97 &    879.62 &  22.342 &  0.026 &   0.741 &  0.061   \\ 
   2985.42 &   2632.81 &  20.519 &  0.017 &  19.348 &  0.020 &    671.49 &    597.82 &  20.891 &  0.008 &   1.443 &  0.012   \\ 
   3739.75 &   2644.34 &  21.703 &  0.014 &  20.755 &  0.019 &    691.16 &    486.85 &  22.119 &  0.016 &   1.293 &  0.027   \\ 
   2296.51 &   2698.13 &  20.280 &  0.015 &  18.652 &  0.028 &    664.63 &    700.93 &  20.853 &  0.009 &   2.068 &  0.013   \\ 
   3216.61 &   2721.97 &  21.870 &  0.010 &  21.008 &  0.010 &    690.23 &    565.81 &  21.594 &  0.012 &   0.869 &  0.026   \\ 
    976.82 &   2729.66 &  20.458 &  0.013 &  19.653 &  0.013 &    637.81 &    896.15 &  20.722 &  0.007 &   0.991 &  0.012   \\ 
   1405.38 &   2776.88 &  20.550 &  0.011 &  19.728 &  0.011 &    655.03 &    834.17 &  20.878 &  0.009 &   1.026 &  0.014   \\ 
\hline
\end{tabular}
}
\label{EMMI_ACS.tab}
\end{table}

\addtocounter{table}{-1}
%\tablenum{1}
\begin{table}
\caption[]{continue.}
\centerline{
\begin{tabular}{|rrrrrr|rrrrrr|}
\hline
\multicolumn{6}{|c|}{ACS} & \multicolumn{6}{c|}{EMMI}\\
\multicolumn{1}{|c}{x}&  \multicolumn{1}{c}{y} & \multicolumn{1}{c}{F606W}&\multicolumn{1}{c}{$\sigma_{F606W}$} &
\multicolumn{1}{c}{F814W}&\multicolumn{1}{c}{$\sigma_{F814W}$} &  \multicolumn{1}{|c}{x}&  \multicolumn{1}{c}{y}& 
\multicolumn{1}{c}{V}&\multicolumn{1}{c}{$\sigma_{V}$} & \multicolumn{1}{c}{I}&\multicolumn{1}{c|}{$\sigma_{I}$} \\
\multicolumn{1}{|c}{pix}& \multicolumn{1}{c}{pix}& \multicolumn{1}{c}{mag}&\multicolumn{1}{c}{mag}&
\multicolumn{1}{c}{mag}&\multicolumn{1}{c}{mag}&\multicolumn{1}{|c}{pix}& \multicolumn{1}{c}{pix}&
\multicolumn{1}{c}{mag}&\multicolumn{1}{c}{mag}&\multicolumn{1}{c}{mag}&\multicolumn{1}{c|}{mag} \\
\hline		   
   3754.16 &   2792.53 &  23.679 &  0.009 &  21.546 &  0.011 &    713.44 &    488.34 &  24.208 &  0.095 &   2.602 &  0.110  \\ 
   2596.22 &   2905.00 &  20.710 &  0.014 &  20.039 &  0.013 &    702.33 &    661.71 &  20.928 &  0.009 &   0.876 &  0.016  \\ 
   3608.48 &   2973.34 &  21.595 &  0.008 &  20.183 &  0.011 &    736.49 &    514.14 &  22.131 &  0.016 &   1.862 &  0.025  \\ 
   3252.45 &   2993.42 &  21.880 &  0.017 &  20.510 &  0.018 &    731.08 &    567.09 &  22.313 &  0.030 &   1.860 &  0.042  \\ 
   1917.10 &   3038.22 &  23.202 &  0.017 &  20.978 &  0.019 &    705.74 &    765.06 &  23.884 &  0.060 &   2.792 &  0.068  \\ 
   3762.43 &   3039.05 &  20.480 &  0.013 &  19.623 &  0.014 &    749.87 &    492.95 &  20.803 &  0.007 &   1.170 &  0.012  \\ 
   3714.12 &   3048.16 &  20.550 &  0.014 &  19.788 &  0.015 &    750.05 &    500.30 &  20.760 &  0.006 &   0.908 &  0.011  \\ 
   2247.26 &   3075.96 &  22.896 &  0.012 &  20.762 &  0.012 &    719.11 &    717.31 &  23.511 &  0.047 &   2.710 &  0.055  \\ 
   2152.90 &   3102.38 &  20.842 &  0.011 &  20.084 &  0.019 &    720.78 &    731.80 &  21.116 &  0.011 &   0.907 &  0.021  \\ 
    557.92 &   3117.79 &  22.646 &  0.014 &  20.694 &  0.015 &    684.85 &    967.12 &  23.405 &  0.050 &   2.579 &  0.057  \\ 
   2906.24 &   3133.20 &  20.304 &  0.017 &  19.455 &  0.019 &    743.31 &    621.48 &  20.565 &  0.005 &   1.033 &  0.011  \\ 
   1693.17 &   3170.59 &  21.351 &  0.017 &  20.280 &  0.019 &    719.84 &    801.17 &  21.793 &  0.012 &   1.404 &  0.021  \\ 
   3829.93 &   3249.78 &  22.354 &  0.013 &  20.255 &  0.013 &    782.62 &    488.05 &  22.956 &  0.032 &   2.721 &  0.037  \\ 
   2369.70 &   3261.08 &  23.113 &  0.020 &  22.367 &  0.020 &    749.36 &    703.61 &  23.395 &  0.055 &   0.754 &  0.129  \\ 
   2314.35 &   3275.75 &  20.546 &  0.012 &  19.738 &  0.014 &    750.18 &    712.14 &  20.852 &  0.007 &   1.003 &  0.012  \\ 
   3169.78 &   3277.28 &  21.489 &  0.016 &  20.577 &  0.017 &    770.82 &    586.10 &  21.832 &  0.021 &   1.179 &  0.034  \\ 
   1772.95 &   3410.03 &  22.544 &  0.014 &  21.649 &  0.014 &    756.90 &    795.12 &  22.838 &  0.031 &   1.106 &  0.060  \\ 
   1942.42 &   3436.93 &  22.119 &  0.017 &  20.199 &  0.015 &    764.96 &    770.82 &  22.763 &  0.031 &   2.556 &  0.039  \\ 
   1571.33 &   3439.63 &  22.400 &  0.016 &  20.925 &  0.016 &    756.49 &    825.59 &  22.902 &  0.031 &   1.929 &  0.042  \\ 
   1970.27 &   3511.46 &  19.634 &  0.025 &  18.865 &  0.025 &    776.71 &    768.51 &  19.869 &  0.005 &   0.823 &  0.009  \\ 
   2059.34 &   3555.50 &  20.410 &  0.014 &  19.397 &  0.014 &    785.26 &    756.44 &  20.840 &  0.007 &   1.328 &  0.012  \\ 
   1586.23 &   3573.68 &  22.616 &  0.013 &  20.849 &  0.012 &    776.70 &    826.61 &  23.225 &  0.045 &   2.297 &  0.052  \\ 
   3558.91 &   3613.37 &  22.513 &  0.010 &  22.417 &  0.016 &    829.65 &    536.79 &  22.586 &  0.021 &   0.282 &  0.086  \\ 
   3498.78 &   3687.52 &  20.827 &  0.010 &  19.348 &  0.025 &    839.05 &    547.34 &  21.322 &  0.009 &   1.968 &  0.013  \\ 
   2483.60 &   3769.17 &  21.306 &  0.014 &  19.866 &  0.012 &    826.89 &    698.98 &  21.841 &  0.013 &   1.936 &  0.019  \\ 
   3942.52 &   3787.70 &  22.105 &  0.009 &  21.235 &  0.011 &    864.41 &    484.30 &  22.474 &  0.023 &   1.332 &  0.038  \\ 
   2806.78 &   3843.25 &  20.498 &  0.016 &  19.556 &  0.019 &    845.49 &    653.14 &  20.816 &  0.007 &   1.230 &  0.014  \\ 
   2673.56 &   4046.14 &  21.995 &  0.014 &  20.015 &  0.013 &    872.12 &    677.60 &  22.667 &  0.024 &   2.625 &  0.029  \\ 
   3180.09 &   4101.70 &  20.500 &  0.018 &  19.677 &  0.022 &    892.49 &    604.25 &  20.806 &  0.006 &   1.116 &  0.011  \\ 
    918.68 &   1230.76 &  19.951 &  0.024 &  18.485 &  0.028 &    415.36 &    868.50 &  20.510 &  0.006 &   1.971 &  0.008  \\ 
   2188.31 &   2811.40 &  20.267 &  0.017 &  19.097 &  0.023 &    678.75 &    719.58 &  20.740 &  0.007 &   1.582 &  0.012  \\ 
   2116.95 &   2905.26 &  20.215 &  0.016 &  19.447 &  0.017 &    690.93 &    732.34 &  20.411 &  0.007 &   0.911 &  0.012  \\ 
   3552.66 &   3237.15 &  20.936 &  0.010 &  19.359 &  0.020 &    774.04 &    528.65 &  21.477 &  0.010 &   2.129 &  0.013  \\ 
   1786.06 &	105.93 &  19.860 &  0.019 &  18.797 &  0.028 &    270.33 &    713.50 &  20.317 &  0.006 &   1.260 &  0.009  \\ 
\hline
\end{tabular}
}
\end{table}

\begin{table}
\caption[]{Average metallicities and spread in [M/H] measured for the NGC~5128 
	halo stars at various distances from the center.}
\centerline{
\begin{tabular}{|ccc|}
\hline
Distance & $\langle$[M/H]$\rangle$ & $\sigma$\\
(kpc)    &     (dex)               &   (dex)  \\
\hline
8        & $-0.46$  & $0.48$ \\
21 - 31  & $-0.69$  & $0.49$ \\
38       & $-0.65$  & $0.49$ \\
\hline
\end{tabular}
}
\label{mean_MH.tab}
\end{table}

\begin{table}
\caption[]{Observed I-band luminosity function of NGC~5128 halo stars 38~kpc away
from the center of the galaxy. $I$ is the magnitude of the bin center,
$N$ are the raw counts in a given magnitude bin 
and $N_{corr}$ are completeness corrected counts.}
\centerline{
\begin{tabular}{|lll|lll|lll|lll|lll|}
\hline
I &  N & N$_{corr}$ & I &  N & N$_{corr}$ &I &  N & N$_{corr}$ &I &  N &
N$_{corr}$ &I &  N & N$_{corr}$ \\
\hline
22.62 & 6  & 6  &   23.88 & 10 & 10    & 25.12 & 92  &  94  &   26.38 & 309 & 326  &	27.62 & 1451 & 1706 \\
22.67 & 8  & 8  &   23.92 & 9  & 9     & 25.17 & 115 & 118  &   26.42 & 295 & 312  &	27.67 & 1855 & 2200 \\
22.73 & 7  & 7  &   23.98 & 12 & 12    & 25.22 & 128 & 131  &   26.47 & 369 & 391  &	27.72 & 2176 & 2606 \\
22.77 & 4  & 4  &   24.02 & 29 & 29    & 25.27 & 132 & 136  &   26.52 & 342 & 363  &	27.77 & 2467 & 2986 \\
22.82 & 11 & 11 &   24.07 & 24 & 24    & 25.32 & 129 & 133  &   26.57 & 346 & 368  &	27.82 & 2840 & 3476 \\
22.88 & 6  & 6	&   24.12 & 45 & 46    & 25.38 & 130 & 134  &   26.62 & 443 & 473  &	27.88 & 2964 & 3672 \\
22.92 & 7  & 7	&   24.17 & 31 & 32    & 25.42 & 147 & 151  &   26.67 & 435 & 465  &	27.92 & 2933 & 3681 \\
22.98 & 1  & 1	&   24.23 & 40 & 41    & 25.47 & 149 & 154  &   26.72 & 503 & 539  &	27.97 & 2811 & 3577 \\
23.02 & 8  & 8	&   24.27 & 55 & 56    & 25.52 & 160 & 165  &   26.77 & 502 & 540  &	28.02 & 2517 & 3251 \\
23.07 & 6  & 6	&   24.32 & 47 & 48    & 25.57 & 138 & 143  &   26.82 & 525 & 566  &	28.07 & 2270 & 2979 \\
23.12 & 7  & 7	&   24.38 & 52 & 53    & 25.62 & 150 & 155  &   26.88 & 525 & 568  &	28.12 & 2006 & 2679 \\
23.17 & 9  & 9	&   24.42 & 63 & 64    & 25.67 & 187 & 194  &   26.92 & 478 & 519  &	28.17 & 1661 & 2259 \\
23.23 & 4  & 4	&   24.48 & 70 & 71    & 25.72 & 173 & 179  &   26.97 & 453 & 494  &	28.22 & 1492 & 2070 \\
23.27 & 8  & 8	&   24.52 & 73 & 74    & 25.77 & 172 & 178  &   27.02 & 387 & 423  &	28.27 & 1475 & 2091 \\
23.32 & 8  & 8	&   24.57 & 98 & 100   & 25.82 & 172 & 179  &   27.07 & 407 & 447  &	28.32 & 1348 & 1955 \\
23.38 & 4  & 4	&   24.62 & 86 & 88    & 25.88 & 217 & 226  &   27.12 & 395 & 436  &	28.38 & 1207 & 1793 \\
23.42 & 8  & 8	&   24.67 & 91 & 93    & 25.92 & 176 & 183  &   27.17 & 504 & 559  &	28.42 & 1272 & 1940 \\
23.48 & 6  & 6	&   24.73 & 99 & 101   & 25.97 & 214 & 223  &   27.22 & 487 & 543  &	28.47 & 1263 & 1980 \\
23.52 & 9  & 9	&   24.77 & 96 & 98    & 26.02 & 206 & 215  &   27.27 & 549 & 615  &	28.52 & 1211 & 1955 \\
23.57 & 6  & 6	&   24.82 & 113 & 116  & 26.07 & 256 & 267  &   27.32 & 610 & 687  &	28.57 & 1252 & 2085 \\
23.62 & 8  & 8	&   24.88 & 115 & 118  & 26.12 & 256 & 268  &   27.38 & 739 & 837  &	28.62 & 1158 & 1993 \\
23.67 & 4  & 4	&   24.92 & 103 & 105  & 26.17 & 233 & 244  &   27.42 & 770 & 878  &	28.67 & 1226 & 2185 \\
23.73 & 7  & 7	&   24.98 & 114 & 117  & 26.22 & 272 & 286  &   27.47 & 830 & 953  &	28.72 & 1178 & 2178 \\
23.77 & 10 & 10 &   25.02 & 115 & 118  & 26.27 & 278 & 292  &   27.52 & 965  & 1116&	28.77 & 1190 & 2286 \\
23.82 & 11 & 11 &   25.07 & 116 & 119  & 26.32 & 274 & 289  &   27.57 & 1194 & 1392&	28.82 & 1250 & 2500 \\
\hline
\end{tabular}
}
\label{N5128LF.tab}
\end{table}

%\clearpage

%\clearpage

\clearpage

%\clearpage

%\clearpage

%\clearpage

\clearpage

\end{document}